\documentclass{llncs}
\usepackage{tikz}

\begin{document}
\mainmatter              
\title{Sharing analysis in the Pawns compiler}
\titlerunning{Pawns sharing}  
\author{Lee Naish}
\authorrunning{L. Naish}   
\tocauthor{Lee Naish}
\institute{Computing and Information Systems,\\
University of Melbourne, Melbourne 3010, Australia\\
\email{dr.lee.naish@gmail.com},\\
\texttt{https://lee-naish.github.io/} }

\maketitle              

\begin{center}
\end{center}
\begin{abstract}

Pawns is a programming language under development that supports algebraic
data types, polymorphism, higher order functions and ``pure'' declarative
programming.  It also supports impure imperative features including
destructive update of shared data structures via pointers, allowing
significantly increased efficiency for some operations.  A novelty of
Pawns is that all impure ``effects'' must be made obvious in the source
code and they can be safely encapsulated in pure
functions in a way that is checked by the compiler.  Execution of a pure
function can perform destructive updates on data structures that are local
to or eventually returned from the function without risking modification
of the data structures passed to the function.  This paper describes the
sharing analysis which allows impurity to be encapsulated.  Aspects of the
analysis are similar to other published work, but in addition it handles
explicit pointers and destructive update, higher order functions including
closures and pre- and post-conditions concerning sharing for functions.

\end{abstract}

\section{Introduction}
\label{sec-intro}

This paper describes the sharing analysis done by the compiler for Pawns
\cite{pawns-overview,pawns}, a programming language that is currently under development.
It is a slightly updated version of \cite{pawns-sharing}, with a section
that briefly describes a new abstract domain used for the analysis,
implemented since the original publication.
Pawns supports both declarative and imperative styles of programming.
It supports algebraic data types, polymorphism, higher order programming
and ``pure'' declarative functions, allowing very high level reasoning
about code.  It also allows imperative code, where programmers can
consider the representation of data types, obtain pointers to the
arguments of data constructors and destructively update them.  Such code
requires the programmer to reason at a much lower level and consider
aliasing of pointers and sharing of data structures.  Low level ``impure''
code can be encapsulated within a pure interface and the compiler checks
the purity.  This requires analysis of pointer aliasing and data structure
sharing, to distinguish data structures that are only visible to the
low level code (and are therefore safe to update) from data structures
that are passed in from the high level code (for which update would
violate purity).  The main aim of Pawns is to get the benefits of purity
for most code but still have the ability to write some key components
using an imperative style, which can significantly improve efficiency
(for example, a more than twenty-fold increase in the speed of inserting
an element into a binary search tree).

There are other functional programming languages, such as ML \cite{ml},
Haskell \cite{haskell} and Disciple \cite{Disciple}, that allow
destructive update of shared data structures but do not allow this
impurity to be encapsulated.  In these languages the ability to update
the data structure is connected to its type\footnote{Disciple uses
``region'' information to augment types, with similar
consequences.}.  For a data structure to be
built using destructive update its type must allow destructive update
and any code that uses the data structure can potentially update it
as well.  This prevents simple declarative analysis of the code and can
lead to a proliferation of different versions of a data structure, with
different parts being mutable. For example, there are four different
versions of lists, since both the list elements and the ``spine'' may
(or may not) be mutable, and sixteen different versions of lists of
pairs.  There is often an efficiency penalty as
well, with destructive update requiring an extra level of indirection
in the data structure (an explicit ``reference'' in the type with most
versions of ML and Haskell).  Pawns avoids this inefficiency and separates
mutability from type information, allowing a data structure to be mutable
in some contexts and considered ``pure'' in others.  The main cost from
the programmer perspective is the need to include extra annotations and
information in the source code.  This can also be considered a benefit,
as it provides useful documentation and error checking.  The main
implementation cost is additional analysis done by the compiler, which
is the focus of this paper.

The rest of this paper assumes some familiarity with
Haskell and is structured as follows.
Section \ref{sec-overview} gives a brief overview of the relevant
features of Pawns.  An early pass of the compiler translates Pawns
programs into a simpler ``core'' language; this is
described in Section \ref{sec-core}.
Section \ref{sec-domain} describes the abstract domain originally used for
the sharing analysis algorithm,
Section \ref{sec-new-domain} describes the new abstract domain that is
now used,
Section \ref{sec-algm} defines the algorithm itself and
Section \ref{sec-example} gives an extended example.
Section \ref{sec-disc} briefly discusses precision and efficiency
issues.
Section \ref{sec-related} discusses related work and
Section \ref{sec-conc} concludes.

\section{An overview of Pawns }
\label{sec-overview}

A more detailed introduction to Pawns is given in \cite{pawns}.
Pawns has many similarities with other functional languages.  It supports
algebraic data types with parametric polymorphism, higher order
programming and curried function definitions.  It uses strict evaluation.
In addition, it supports destructive update via ``references'' (pointers)
and has a variety of extra annotations to make impure effects more clear
from the source code and allow them to be encapsulated in pure code.
Pawns also supports a form of global variables (called state variables)
which support encapsulated effects, but we do not discuss them further
here as they are handled in essentially the same way as other variables
in sharing analysis.  Pure code can be thought of in a declarative way,
were values can be viewed abstractly, without considering how they
are represented.  Code that uses destructive update must be viewed at a
lower level, considering the representation of values, including sharing.
We discuss this lower level view first, then briefly present how impurity
can be encapsulated to support the high level view.  We use Haskell-like
syntax for familiarity.

\subsection{The low level view}
\label{sec-lowlev}

Values in Pawns are represented as follows.  Constants (data constructors
with no arguments) are represented using a value in a single word.
A data constructor with $N>0$ arguments is represented using a word
that contains a tagged pointer to a block of $N$ words in main memory
containing the arguments.  For simple data types such as lists the tag
may be empty.  In more complex cases some bits of the pointer may be used
and/or a tag may be stored in a word in main memory along with the arguments.
Note that constants and tagged pointers are not always stored in main
memory and Pawns variables may correspond to registers that contain
the value.  Only the arguments of data constructors are guaranteed to
be in main memory.  An array of size $N$ is represented in the same
way as a data constructor with $N$ arguments, with the size given by
the tag. Functions are represented as either a constant (for functions
that are known statically) or a closure which is a data constructor
with a known function and a number of other arguments.

Pawns has a \texttt{Ref t} type constructor, representing a
reference/pointer to a value of type \texttt{t} (which must be stored
in memory).  Conceptually we can think of a corresponding \texttt{Ref}
data constructor with a single argument, but this is never explicit
in Pawns code.  Instead, there is an explicit dereference operation:
\texttt{*vp} denotes the value \texttt{vp} points to.  There are two
ways references can be created: let bindings and pattern bindings.  A let
binding \texttt{*vp = val} allocates a word in main memory, initializes it
to \texttt{val} and makes \texttt{vp} a reference to it (Pawns omits
Haskell's \texttt{let} and \texttt{in} keywords; the scope is the
following sequence of statements/expressions).  In a pattern
binding, if \texttt{*vp} is the argument of a data constructor pattern,
\texttt{vp} is bound to a reference to the corresponding argument of the
data constructor if pattern matching succeeds (there is also a primitive
that returns a reference to the $i^{th}$ element of an array).
Note it is not possible
to obtain a reference to a Pawns variable: variables do not denote memory
locations.  However, a variable \texttt{vp} of type \texttt{Ref t} denotes
a reference to a memory location containing a value of type \texttt{t} and
the memory location can be destructively updated by \texttt{*vp := val}.

Consider the following code.  Two data types are defined. The code
creates a reference to \texttt{Nil} (\texttt{Nil} is stored in a newly
allocated memory word) and a reference to that reference (a pointer
to the word containing \texttt{Nil} is put in another allocated word).
It also creates a list containing constants \texttt{Blue} and \texttt{Red}
(requiring the allocation of two cons cells in memory; the \texttt{Nil}
is copied).  It deconstructs the list to obtain pointers to the head and
tail of the list (the two words in the first cons cell) then destructively
updates the head of the list to be \texttt{Red}.

\begin{verbatim}
data Colour = Red | Green | Blue
data Colours = Nil | Cons Colour Colours -- like [Colour]
\end{verbatim}
\begin{verbatim}
    ...
    *np = Nil                       -- np = ref to (copy of) Nil
    *npp = np                       -- npp = ref to (copy of) np
    cols = Cons Blue (Cons Red *np) -- cols = [Blue, Red]
    case cols of
    (Cons *headp *tailp) ->         -- get ref to head and tail
        *headp := Red               -- update head with Red
\end{verbatim}

The memory layout after the assignment can be pictured as follows,
where boxes represent main memory words and \texttt{Ref} and \texttt{Cons}
followed by an arrow represent pointers (no tag is used in either case):

\noindent
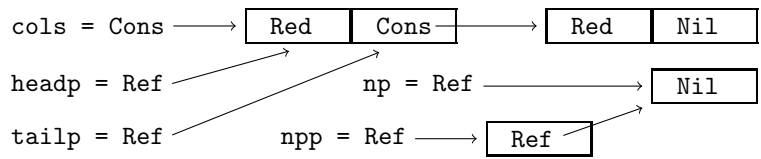
\begin{tikzpicture}
\begin{scope}[scale=1.0]
\node at (0,1.5) (c1) {\texttt{cols = Cons}};
\node at (3.5,1.5) (c2)
        {\framebox[4em]{\texttt{Red }}\framebox[4em]{\texttt{Cons}}};
\node at (7.5,1.5) (ct)
        {\framebox[4em]{\texttt{Red }}\framebox[4em]{\texttt{Nil }}};
\draw[->] (c1) -- (c2);
\node at (2.8,1.2) (h) {};
\node at (4.5,1.5) (t1) {};
\node at (4.0,1.2) (t) {};
\node at (8.2,0.7) (nb)
        {\framebox[4em]{\texttt{Nil }}};
\node at (4.4,0.7) (np) {\texttt{np = Ref}};
\node at (3.4,0) (npp) {\texttt{npp = Ref}};
\node at (6.0,0) (npb)
        {\framebox[4em]{\texttt{Ref }}};
\node at (6.2,0.0) (r) {};
\draw[->] (npp) -- (npb);
\draw[->] (r) -- (nb);
\draw[->] (np) -- (nb);
\node at (0,0.7) (hp) {\texttt{headp = Ref}};
\node at (1,0.7) (hp1) {};
\node at (0,0) (tp) {\texttt{tailp = Ref}};
\node at (1,0) (tp1) {};
\draw[->] (hp1) -- (h);
\draw[->] (t1) -- (ct);
\draw[->] (tp1) -- (t);
\end{scope}
\end{tikzpicture}

The destructive update above changes the values of both \texttt{headp} and
\texttt{cols} (the representations are shared).  One of the novel features
of Pawns is that the source code must be annotated with ``!'' to make it
obvious when each ``live'' variable is updated.  If both \texttt{headp}
and \texttt{cols} are used later, the assignment statement above must
be written as follows, with \texttt{headp} prefixed with ``!'' and an
additional annotation attached to the whole statement indicating
\texttt{cols} may be updated:

\begin{verbatim}
        *!headp := Red  !cols        -- update *headp (and cols)
\end{verbatim}

We say that the statement \emph{directly} updates \texttt{headp}
and \emph{indirectly} updates \texttt{cols}, due to sharing of
representations.  Similarly, if \texttt{headp} was passed to a
function that may update it, additional annotations are required.
For example, \texttt{(assign !headp Red) !cols} makes the direct
update of \texttt{headp} and indirect update of \texttt{cols} clear.
Sharing analysis is used to ensure that source code contains all the
necessary annotations.  One aim of Pawns is that any effects of code
should be made clear by the code.  Pawns is an acronym for Pointer
Assignment With No Surprises.

Pawns functions have extra annotations in type signatures to
document which arguments may be updated.  For additional documentation,
and help in sharing analysis, there are annotations to declare what
sharing may exist between arguments when the function is called
(a precondition) and what extra sharing may be added by executing the
function (called a postcondition, though it is the union of the pre- and
post-condition that must be satisfied after a function is executed).
For example, we may have:
\pagebreak[4]
\begin{verbatim}
assign :: Ref t -> t -> ()
    sharing assign !p v = _   -- p may be updated
    pre nosharing             -- p&v don't share when called
    post *p = v               -- assign may make *p alias with v
assign !p v =
    *!p := v
\end{verbatim}

The ``!'' annotation on parameter \texttt{p} declares the first
argument of \texttt{assign} is mutable.  The default is that arguments
are not mutable.
As well as checking for annotations on assignments and function calls,
sharing analysis is used to check that all parameters which may be
updated are declared mutable in type signatures, and pre- and post-conditions
are always satisfied.  For example, assuming the previous code which
binds \texttt{cols}, the call \texttt{assign !tailp !cols} annotates all
modified variables but violates the precondition of \texttt{assign}
because there is sharing between \texttt{tailp} and \texttt{cols} at the
time of the call.  Violating this precondition allows cyclic structures
to be created, which is important for understanding the code.
If the precondition was dropped, the second argument of \texttt{assign}
would also need to be declared mutable in the type signature and the
assignment to \texttt{p} would require \texttt{v} to be annotated.
In general, there is an inter-dependence between ``!'' annotations in the
code and pre- and post-conditions.  More possible sharing at a call means
more ``!'' annotations are needed, more sharing in (recursive) calls and
more sharing when the function returns.

Curried functions and higher order code are supported by attaching
sharing and destructive update information to each arrow in a type,
though often the information is inferred rather than being given
explicitly in the source code.  For example, implicit in the declaration
for \texttt{assign} above is that \texttt{assign} called with a single
argument of type \texttt{Ref t} creates a closure of type \verb@t -> ()@
containing that argument (and thus sharing the object of type
\texttt{t}).  The explicit sharing information describes applications
of this closure to another argument.  There is a single argument in
this application, referred to with the formal parameter \texttt{v}.
The other formal parameter, \texttt{p}, refers to the argument of the
closure.  In general, a type with $N$ arrows in the ``spine'' has $K+N$
formal parameters in the description of sharing, with the first $K$
parameters being closure arguments.

The following code defines binary search trees of integers
and defines a function that takes a pointer to a tree and
inserts an integer into the tree.  It uses destructive update, as would
normally be done in an imperative language.  The declarative alternative
must reconstruct all nodes in the path from the root down to the new node.
Experiments using our prototype implementation of Pawns indicate that for
long paths this destructive update version is as fast as hand-written
C code whereas the ``pure'' version is more than twenty times slower,
primarily due to the overhead of memory allocation.

\pagebreak[4]
\begin{verbatim}
data Tree = TNil | Node Tree Int Tree
\end{verbatim}
\begin{verbatim}
bst_insert_du :: Int -> Ref Tree -> ()
    sharing bst_insert_du x !tp = _   -- tree gets updated
    pre nosharing                     -- integers are atomic so
    post nosharing                    -- it doesn't share
bst_insert_du x !tp =
    case *tp of
    TNil ->
        *!tp := Node TNil x TNil      -- insert new node
    (Node *lp n *rp) ->
        if x <= n then
            (bst_insert_du x !lp) !tp -- update lp (and tp)
        else
            (bst_insert_du x !rp) !tp -- update rp (and tp)
\end{verbatim}

\subsection{The high level view}
\label{sec-highlev}

Whenever destructive update is used in Pawns, programmers must be aware
of potential sharing of data representations and take a low level view.
In other cases it is desirable to have a high level view of values,
ignoring how they are represented and any sharing that may be present.
For example, in the two trees \texttt{t1} and \texttt{t2} depicted below,
it is much simpler if we do not have to care or know about the sharing
between the trees and within tree \texttt{t1}.  The high level view
is they are both just \texttt{Node (Node TNil 123 TNil) 123 (Node TNil
123 TNil)}.

\noindent
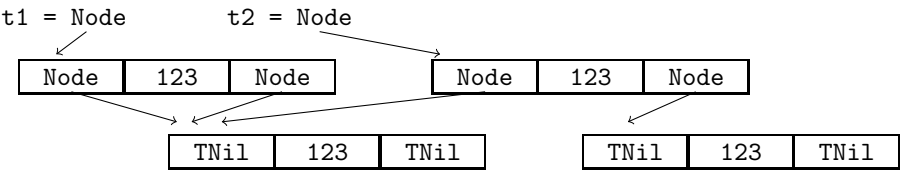
\begin{tikzpicture}
\begin{scope}[scale=1.0]
\node at (-0.5,1.8) (t1) {\texttt{t1 = Node}};
\node at (2.5,1.8) (t2) {\texttt{t2 = Node}};
\node at (1.0,1.0) (t1a)
  {\framebox[4em]{\texttt{Node}}\framebox[4em]{\texttt{123}}\framebox[4em]{\texttt{Node}}};
\draw[->] (-0.2,1.6) -- (-0.6,1.3);
\draw[->] (2.9,1.6) -- (4.5, 1.3);
\node at (6.5,1.0) (t2a)
  {\framebox[4em]{\texttt{Node}}\framebox[4em]{\texttt{123}}\framebox[4em]{\texttt{Node}}};
\node at (3.0,0.0) (l1a)
  {\framebox[4em]{\texttt{TNil}}\framebox[4em]{\texttt{123}}\framebox[4em]{\texttt{TNil}}};
\draw[->] (-0.4,0.8) -- (1.0,0.4);
\draw[->] (2.4,0.8) -- (1.2, 0.4);
\node at (8.5,0.0) (l2a)
  {\framebox[4em]{\texttt{TNil}}\framebox[4em]{\texttt{123}}\framebox[4em]{\texttt{TNil}}};
\draw[->] (5.1,0.8) -- (1.6,0.4);
\draw[->] (7.9,0.8) -- (7.0, 0.4);
\end{scope}
\end{tikzpicture}

Pawns has a mechanism to indicate that the high level view is
taken.  Pre- and post-conditions can specify sharing with a special
pseudo-variable named \texttt{abstract}\footnote{There is conceptually
a different \texttt{abstract} variable for each distinct type.}.
The sharing analysis of the Pawns compiler allows a distinction between
``abstract'' variables, which share with \texttt{abstract} and for which
the programmer takes a high level view, and ``concrete'' variables for
which the programmer must understand the representation and explicitly
declare all sharing in pre- and post-conditions.  The analysis checks
that no live abstract variables can be destructively updated.  Thus if a
function has a parameter which is updated, it must be declared mutable and
must not be declared to share with \texttt{abstract} in the precondition
(non-mutable parameters may or may not share with \texttt{abstract}).
Checking of preconditions ensures that abstract variables are not
passed to functions which expect concrete data structures.  For example,
an abstract tree cannot be passed to \verb@bst_insert_du@ because the
precondition allows no sharing with \texttt{abstract}.  It is important
that the tree structure is known when \verb@bst_insert_du@ is used
because the result depends on it.  For example, inserting into the right
subtree of \texttt{t2} only affects this subtree whereas inserting into
the right subtree of \texttt{t1} (which has the same high level value)
also changes the left subtree of both \texttt{t1} and \texttt{t2}.
Note that concrete variables can be passed to functions which allow
abstract arguments.  Pawns type
signatures that have no annotations concerning destructive update or
sharing implicitly indicate no arguments are destructively updated and
the arguments and result share with \texttt{abstract}. Thus a subset of
Pawns code can look like and be considered as pure functional code.

The following code defines a function that takes a list of integers and
returns a binary search tree containing the same integers.  Though it uses
destructive update internally, this impurity is encapsulated and it can
therefore be viewed as a pure function.  The list that is passed in as an
argument is never updated and the tree returned is abstract so it is
never subsequently updated (a concrete tree could be returned if an
explicit postcondition without \texttt{t = abstract} was given).
An initially empty tree is created locally.
It is destructively updated by inserting each integer of the list into it
(using \verb@list_bst_du@, which calls \verb@bst_insert_du@), then the
tree is returned.  Within the execution of \verb@list_bst@ it is important
to understand the low level details of how the tree is represented,
but this information is not needed outside the call.

\begin{verbatim}
data Ints = Nil | Cons Int Ints

list_bst :: Ints -> Tree  -- pure function from Ints to Tree
 -- implicit sharing information:
 -- sharing list_bst xs = t
 -- pre xs = abstract
 -- post t = abstract
list_bst xs =
    *tp = TNil            -- create pointer to empty tree
    list_bst_du xs !tp    -- insert integers into tree
    *tp                   -- return (updated) tree
\end{verbatim}
\begin{verbatim}
list_bst_du :: Ints -> Ref Tree -> ()
    sharing list_bst_du xs !tp = _  -- tree gets updated
    pre xs = abstract
    post nosharing
list_bst_du xs !tp =
    case xs of
    (Cons x xs1) ->
       bst_insert_du x !tp  -- insert head of list into tree
       list_bst_du xs1 !tp  -- insert rest of list into tree
    Nil -> ()
\end{verbatim}

\section{Core Pawns}
\label{sec-core}

An early pass of the Pawns compiler converts all function definitions
into a core language by flattening nested expressions, introducing extra
variables et cetera.  A variable representing the return value of the
function is introduced and expressions are converted to bindings for
variables.  A representation of the core language version of code is
annotated with type, liveness and other information prior to sharing
analysis.  We just describe the core language here.  The right side of
each function definition is a statement (described using the definition
of type \texttt{Stat} below), which may contain variables, including
function names (\texttt{Var}), data constructors (\texttt{DCons})
and pairs containing a pattern (\texttt{Pat}) and statement for case
statements.  All variables are distinct except for those in recursive
instances of \texttt{Stat} and variables are renamed to avoid any
ambiguity due to scope.

\begin{verbatim}
data Stat =                   -- Statement, eg
    Seq Stat Stat |           -- stat1 ; stat2
    EqVar Var Var |           -- v = v1
    EqDeref Var Var |         -- v = *v1
    DerefEq Var Var |         -- *v = v1
    DC Var DCons [Var] |      -- v = Cons v1 v2
    Case Var [(Pat, Stat)] |  -- case v of pat1 -> stat1 ...
    Error |                   -- (for uncovered cases)
    App Var Var [Var] |       -- v = f v1 v2
    Assign Var Var |          -- *!v := v1
    Instype Var Var           -- v = v1::instance_of_v1_type

data Pat =                    -- patterns for case, eg
    Pat DCons [Var]           -- (Cons *v1 *v2)
\end{verbatim}

Patterns in the core language only bind references to arguments --- the
arguments themselves must be obtained by explicit dereference operations.
Pawns supports ``default'' patterns but for simplicity of presentation
here we assume all patterns are covered in core Pawns and we include an
error primitive.  Similarly, we just give the general case for application
of a variable to $N>0$ arguments; our implementation distinguishes some
special cases.  Memory is allocated for \texttt{DerefEq}, \texttt{DC}
(for non-constants) and \texttt{App} (for unsaturated applications which
result in a closure).
The runtime behaviour of \texttt{Instype} is identical to \texttt{EqVar}
but it is treated differently in type analysis.

Sharing and type analysis
cannot be entirely separated.  Destructive update in the presence of
polymorphic types can potentially violate type safety or ``preservation''
--- see \cite{mlvalrest}, for example.  For a variable whose type is
polymorphic (contains a type variable), we must avoid assigning a value
with a less general type.  For example, in \texttt{*x = []} the type
of \texttt{*x} is ``list of \texttt{t}'', where \texttt{t} is a type variable.
Without destructive update it should be possible to use \texttt{*x}
wherever a list of any type is expected.  However, if \texttt{*x} is
then assigned a list containing integers (which has a less general type),
passing it to a function that
expects a list of functions violates type safety (``calling'' an arbitrary
integer is not safe).  Pawns allows expressions to have their inferred
types further instantiated using ``::'', and the type checking pass of
the compiler also inserts some type instantiation.  The type checking
pass ensures that direct update does not involve type instantiation but
to improve flexibility, indirect update is checked during the sharing
analysis.

\section{The abstract domain}
\label{sec-domain}

The representation of the value of a variable includes some set of main
memory words (arguments of data constructors).  Two variables share if
the intersection of their sets of main memory words is not empty. The
abstract domain for sharing analysis must maintain a conservative
approximation to all sharing, so we can tell if two variables possibly
share (or definitely do not share).  The abstract domain we use is a
set of pairs (representing possibly intersecting sets of main memory
locations) of variable \emph{components}.  The different components of
a variable partition the set of main memory words for the variable.

The components of a variable depend on its type.  For non-recursive
types other than arrays, each possible data constructor argument is
represented separately.  For example,
the type \texttt{Maybe (Maybe (Either Int Int))} can have an argument
of an outer \texttt{Just} data constructor, an inner \texttt{Just}
and \texttt{Left} and \texttt{Right}.  A component can be represented
using a list of \verb@x.y@ pairs containing a data constructor and an
argument number, giving the path from the outermost data constructor
to the given argument.  For example, the components of the type above
can be written as: \verb@[Just.1]@, \verb@[Just.1,Just.1]@,
\verb@[Just.1,Just.1,Left.1]@ and \verb@[Just.1,Just.1,Right.1]@.
If variable \texttt{v} has value \texttt{Just Nothing}, the expression
\verb@v.[Just.1]@ represents the single main memory word containing the
occurrence of \texttt{Nothing}.

For \texttt{Ref t} types we proceed as if there was a \texttt{Ref}
data constructor, so \verb@vp.[Ref.1]@ represents the word \texttt{vp}
points to.  For function types, values may be closures.  A closure that
has had $K$ arguments supplied is represented as a data constructor
$\mathtt{Cl}_K$ with these $K$ arguments; these behave in the same way as
other data constructor arguments with respect to sharing,
except Pawns provides no way to obtain a pointer to a closure argument.
Closures also
contain a code pointer and an integer which are not relevant to sharing so
they are ignored in the analysis.
We also ignore the subscript on the data constructor
for sharing analysis because type and sharing analysis only give a
lower bound on the number of closure arguments.  Our analysis orders
closure arguments so that the most recently supplied argument is first
(the reverse of the more natural ordering).  Consider the code below,
where \texttt{foo} is a function that is defined with four or more
arguments.  The sharing analysis proceeds as if the memory layout was as
depicted in the diagram.  The pre- and post-conditions of \texttt{foo}
are part of the type information associated with \texttt{c1},
\texttt{c2} and \texttt{c3}.

\noindent
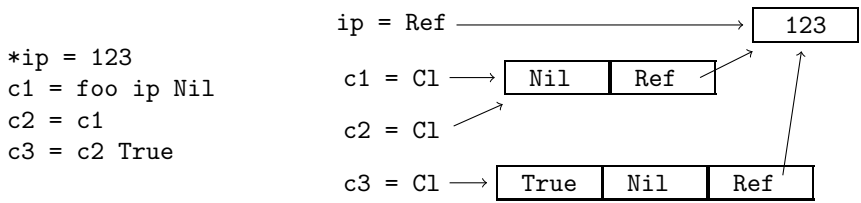
\begin{minipage}[c]{0.4\textwidth}
\begin{verbatim}
    *ip = 123
    c1 = foo ip Nil
    c2 = c1
    c3 = c2 True
\end{verbatim}
\end{minipage}
\begin{minipage}[c]{0.5\textwidth}
\noindent
\begin{tikzpicture}
\begin{scope}[scale=1.0]
\node at (0,3.3) (ip1) {\texttt{ip = Ref}};
\node at (5.5,3.3) (ip2)
        {\framebox[4em]{\texttt{ 123 }}};
\draw[->] (ip1) -- (ip2);
\node at (0,2.6) (c1) {\texttt{c1 = Cl}};
\node at (2.9,2.6) (c1a)
        {\framebox[4em]{\texttt{Nil }}\framebox[4em]{\texttt{Ref  }}};
\draw[->] (c1) -- (c1a);
\node at (2.8,2.6) (c1b) {};
\node at (0,1.9) (c2) {\texttt{c2 = Cl}};
\node at (0.7,1.9) (c2a) {};
\node at (1.6,2.3) (c2b) {};
\draw[->] (c2a) -- (c2b);
\draw[->] (4.1,2.6) -- (ip2) {};
\node at (0,1.2) (c3) {\texttt{c3 = Cl}};
\node at (3.5,1.2) (f2)
        {\framebox[4em]{\texttt{True}}\framebox[4em]{\texttt{Nil
}}\framebox[4em]{\texttt{Ref  }}};
\node at (2.8,1.2) (f2a) {};
\draw[->] (c3) -- (f2);
\draw[->] (5.2,1.3) -- (ip2) {};
\node at (2.8,1.4) (f2c) {};
\end{scope}
\end{tikzpicture}
\end{minipage}

For arrays, \verb@[Array_.1]@ is used to represent all words in the array.
The expression, \verb@x.[Array_.1,Just.1]@ represents the arguments of all
\texttt{Just} elements in an array \texttt{x} of \texttt{Maybe} values.
For recursive types, paths are ``folded'' \cite{type_fold}
so there are a finite number
of components.  Here we present the original published version of the
abstract domain, which is also used in later examples of the sharing
analysis algorithm.  The compiler now uses a more expressive abstract
domain for recursive types, described in Section \ref{sec-new-domain}.
If a type $T$ has sub-component(s) of type $T$ we use
the empty path to denote the sub-component(s).  In general, we construct
a path from the top level and if we come across a sub-component of type $T$
that is in the list of ancestor types (the top level type followed by the
types of elements of the path constructed so far) we just use the path
to the ancestor to represent the sub-component.  Consider the following
mutually recursive types that can be used to represent trees which
consist of a node containing an integer and a list of sub-trees:

\begin{verbatim}
data RTrees = Nil | Cons RTree RTrees
data RTree = RNode Int RTrees
\end{verbatim}

For type \texttt{RTrees} we have the components
\verb@[]@ (this folded path represents both \verb@[Cons.2]@ and
\verb@[Cons.1,RNode.2]@, since they are of type \texttt{RTrees}),
\verb@[Cons.1]@ and
\verb@[Cons.1,RNode.1]@.
The expression \verb@t.[Cons.1,RNode.1]@ represents the set of memory
words that
are the first argument of \texttt{RNode} in variable \texttt{t} of type
\texttt{RTrees}. 
For type \texttt{RTree} we have the components
\verb@[]@ (for \verb@[RNode.2,Cons.1]@, of type \texttt{RTree}),
\verb@[RNode.1]@ and
\verb@[RNode.2]@ (which is also the folded version of
\verb@[RNode.2,Cons.2]@, of type \texttt{RTrees}).
In our sharing analysis algorithm we use a function \texttt{fc} (fold
component) which takes a $v.c$ pair,
and returns $v.c'$ where $c'$ is the correctly folded component for the
type of variable $v$.  For example, \verb@fc (ts.[Cons.2])@ =
\verb@ts.[]@, assuming \texttt{ts} has type \texttt{RTrees}.

As well as containing pairs of components for distinct variables which may
alias, the abstract domain contains ``self-alias'' pairs for
each possible component of a variable which may exist.  Consider the
following two bindings and the corresponding diagram (as with
\texttt{Cons}, no tag is used for \texttt{RNode}):

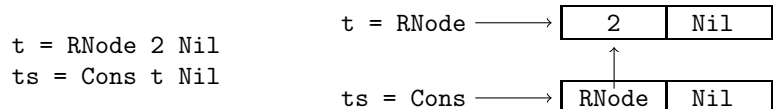
\begin{minipage}[c]{0.4\textwidth}
\begin{verbatim}
    t = RNode 2 Nil
    ts = Cons t Nil
\end{verbatim}
\end{minipage}
\begin{minipage}[c]{0.5\textwidth}
\noindent
\begin{tikzpicture}
\begin{scope}[scale=1.0]
\node at (0,1.7) (c1) {\texttt{t = RNode}};
\node at (3.5,1.7) (c2)
        {\framebox[4em]{\texttt{ 2   }}\framebox[4em]{\texttt{Nil  }}};
\draw[->] (c1) -- (c2);
\node at (0,0.7) (d1) {\texttt{ts = Cons}};
\node at (3.5,0.7) (d2)
        {\framebox[4em]{\texttt{RNode}}\framebox[4em]{\texttt{Nil  }}};
\node at (2.8,1.5) (c2a) {};
\node at (2.8,0.7) (d2a) {};
\draw[->] (d1) -- (d2);
\draw[->] (d2a) -- (c2a);
\end{scope}
\end{tikzpicture}
\end{minipage}

With our domain, the most precise description of sharing after
these two bindings is as follows.  We represent an alias pair as a set
of two variable components.  The first five are self-alias pairs and
the other two describe the sharing between \texttt{t} and \texttt{ts}.

\pagebreak[4]
\begin{verbatim}
   {{t.[RNode.1], t.[RNode.1]},
    {t.[RNode.2], t.[RNode.2]},
    {ts.[], ts.[]},
    {ts.[Cons.1], ts.[Cons.1]},
    {ts.[Cons.1,RNode.1], ts.[Cons.1,RNode.1]},
    {t.[RNode.1], ts.[Cons.1,RNode.1]},
    {t.[RNode.2], ts.[]}}
\end{verbatim}	

Note there is no self-alias pair for \verb@t.[]@ since there is no strict
sub-part of \texttt{t} that is an \texttt{RTree}.  Similarly, there
is no alias between \verb@ts.[Cons.1]@ and any part of \texttt{t}.
Although the value \texttt{t} is used as the first argument of
\texttt{Cons} in \texttt{ts}, this is not a main memory word that is
used to represent the value of \texttt{t} (indeed, the value of \texttt{t}
has no \texttt{Cons} cells).  The tagged pointer value stored in variable
\texttt{t} (which may be in a register) is copied into the cons cell. 
Such descriptions of sharing are an abstraction of computation states.
The set above abstracts all computation states in which \texttt{t}
is a tree with a single node, \texttt{ts} is a list of trees, elements
of \texttt{ts} may be \texttt{t} or have \texttt{t} as a subtree, and
there are no other live variables with non-atomic values.

\section{The new abstract domain}
\label{sec-new-domain}

The Pawns compiler no longer uses empty paths to represent components
of variables with recursive types.  One of the motivations was to get
more precise sharing analysis for code such as the following, which
repeatedly assigns different RTrees to a single pointer variable
(similar imprecision occured for other recursively defined types).
\begin{verbatim}
*tp = rtree1; ... *!tp := rtree2; ... *!tp := rtree3;
\end{verbatim}
With the original domain, the memory word \verb@*tp@ is represented
by \verb@tp.[Ref.1]@, which also represents words that are the first
argument of a \texttt{Cons} cells within the tree.  The compiler
concludes the assignments may update \verb@rtree1@ and \verb@rtree2@ and
introduce sharing between \verb@rtree1@, \verb@rtree2@ and \verb@rtree3@.
The code can be re-written by renaming \texttt{tp} for each assignment
to get more precised analysis, but it can be inconvenient for programmers
to repeatedly rename variables and ideally it should not be necessary.

The new abstract domain used by the compiler contains all non-empty paths
that avoid repetitions corresponding to recursion in the type. Thus
\verb@rtree1@ will typically have four components: \verb@[RNode.1]@,
\verb@[RNode.2,Cons.1]@ (instead of \verb@[]@), and both \verb@[RNode.2]@
and \verb@[RNode.2,Cons.2]@. Only paths with repeated data constructors
are folded.  For example, \verb@[RNode.2,Cons.2,Cons.2]@ is folded
to become \verb@[RNode.2,Cons.2]@. Note that \verb@[RNode.2]@
represents memory words that arguments of \texttt{RNode} whereas
\verb@[RNode.2,Cons.2]@ represents memory words that arguments of
\texttt{Cons}, and these must be distinct, even though they have the
same type. This distinction was not made in the old domain.  Similarly,
there will typically be five components of \texttt{tp}, \verb@[Ref.1]@
plus the four tree components prefixed by \verb@[Ref.1]@, so there is a
distinction made between the memory word used for \verb@*tp@ and those
used by the tree. With the new abstract domain, sharing analysis for
let bound ref variables loses no information unless there is recursion
through the \texttt{ref} type. Note that the examples below use the old
abstract domain.  Also, the sharing algorithm given below treats empty
paths as special in some places so the current compiler uses a some small
modifications to this algorithm.

\section{The sharing analysis algorithm}
\label{sec-algm}

We now describe the sharing analysis algorithm.  Overall, the compiler
attempts to find a proof that for a computation with a depth $D$ of
(possibly recursive) function calls, the following condition $C$ holds,
assuming $C$ holds for all computations of depth less than $D$.  This
allows a proof by induction that $C$ holds for all computations that
terminate normally.

\vspace{2mm}
\noindent
$C$: For all functions $f$, if the precondition of
$f$ is satisfied (abstracts the computation state)
whenever $f$ is called, then
\begin{enumerate}
\item for all function calls and assignment statements in $f$, any live
variable that may be updated at that point in an execution of $f$ is
annotated with ``!'',
\item there is no update of live ``abstract'' variables when executing
$f$,
\item all parameters of $f$ which may be updated when
executing $f$ are declared mutable in the type signature of $f$,
\item the union of the pre- and post-conditions of $f$ abstracts
the state when $f$ returns plus the values
of mutable
parameters in all states during the execution of $f$,
\item for all function calls and assignment statements in $f$, any live
variable that may be directly updated at that point is updated with a value
of the same type or a more general type, and
\item for all function calls and assignment statements in $f$, any live
variable that may be indirectly updated at that point only shares with
variables of the same type or a more general type.
\end{enumerate}

The algorithm is applied to each function definition
in core Pawns to compute an approximation to the sharing before and
after each statement (we call it the alias set).  This can be used to
check points 1, 2, 4 and 6 above. The algorithm checks that
preconditions are satisfied for each function call, allowing
the induction hypothesis to be used.  Point 3
is established using point 1 and a simple syntactic check that any
parameter of $f$ that is annotated ``!'' in the definition is declared
mutable in the type signature (parameters are considered live throughout
the definition).  Point 5 relies on 3 and the type checking pass.
The core of the algorithm is to compute the alias set
after a statement, given the alias set before the statement.  This is
applied recursively for compound statements in a form of abstract
execution.  Note that for point 4, if a statement changes the set of
memory cells used to represent a mutable parameter, the algorithm computes
the sharing for the union of the two sets of cells.

We do not prove correctness of the algorithm but hope our presentation
is sufficiently detailed to have uncovered any bugs.  A proof would
have a separate case for each kind of statement in the core language,
showing that if the initial alias set abstracts the execution state
before the statement the resulting alias set abstracts the execution
state after the statement.  This would require a more formal description
of execution states and their relationship with the core language and
the abstract domain.  The abstract domain relies on type information so
the sharing analysis relies on type preservation in the execution. Type
preservation also relies on sharing analysis.  Thus a completely formal
approach must tackle both problems together.  Although our approach is not
formal, we do state the key condition $C$, which has points relating to
both sharing and types, and we include \texttt{Instype} in the core language.

The alias set used at the start of a definition is the precondition
of the function.  This implicitly includes self-alias pairs
for all variable components of the arguments of the function and
the pseudo-variables $\mathtt{abstract}_T$ for each type $T$ used.
Similarly, the postcondition implicitly includes self-alias pairs for
all components of the result (and the $\mathtt{abstract}_T$ variable if
the result is abstract)\footnote{Self-aliasing for arguments and results is
usually desired.  For the rare cases it is not, we may provide a mechanism
to override this default in the future.}.  As abstract execution proceeds,
extra variables from the function body are added to the alias set and
variables that are no longer live can be removed to improve efficiency.
For each program point, the computed alias set abstracts the computation
state at that point in all concrete executions of the function that
satisfy the precondition.  For mutable parameters of the function, the
sharing computed also includes the sharing from previous program points.
The reason for this special treatment is explained when we discuss the
analysis of function application.  The alias set computed for the end
of the definition, with sharing for local variables removed, must be a
subset of the union of the pre- and post-condition of the function.

Before sharing analysis, a type checking/inference pass is completed which
assigns a type to each variable and function application.  This
determines the components for each variable.  Polymorphism is also
eliminated as follows.  Suppose we have a function \texttt{take n xs},
which returns the list containing the first \texttt{n} elements of
\texttt{xs}:

\pagebreak[4]
\begin{verbatim}
take :: Int -> [a] -> [a]
    sharing take n xs = ys
    pre nosharing
    post ys = xs
\end{verbatim}
For each call to \texttt{take}, the pre- and post-conditions
are determined based on the type of the application.  An application to
lists of Booleans will have two components for each variable whereas an
application to lists of lists of Booleans will have four.  When analysing
the definition of \texttt{take} we instantiate type variables such as
\texttt{a} above to \texttt{Ref ()}.  This type has a single component
which can be shared to represent possible sharing of arbitrary components
of an arbitrary type.
Type checking prevents sharing between non-identical types, such as
\verb@[a]@ and \verb@[b]@.  Finally, we assume there is no type which is
an infinite chain of refs, for example, \texttt{type Refs = Ref Refs}
(for which type folding results in an empty component rather than a
\verb@[Ref.1]@ component; this is not a practical limitation).

Suppose $a_0$ is the alias set just before statement $s$.  The following
algorithm computes $\mathtt{alias}(s, a_0)$, the alias set just after
statement $s$.  The algorithm structure follows the recursive definition
of statements and we describe it using pseudo-Haskell, interspersed with
discussion. The empty list is written $\texttt{[]}$, non-empty lists
are written $\mathtt{[a,b,c]}$ or $\texttt{a:b:c:[]}$ and $\mathtt{++}$
denotes list concatenation.
At some points we use high level declarative set
comprehensions to describe what is computed and naive implementation may
not lead to the best performance.

\begin{tabbing}
mm\=mm\=mm\=mm\=mm\=mm\=mm\=mm\=mm\=mm\=mm\=mm\=mm\=mm\kill
\texttt{alias (Seq stat1 stat2) a0 = }\>\>\>\>\>\>\>\>\>\>\>\>\>\texttt{-- stat1; stat2}\\
\> \texttt{alias stat2 (alias stat1 a0) }\\
\texttt{alias (EqVar v1 v2) a0 = }\>\>\>\>\>\>\>\>\>\>\>\>\>\texttt{-- v1 = v2}\\
\> \texttt{let} \\
\> \> \texttt{self1 = }
    $\{ \{\mathtt{v1.}c_1,\mathtt{v1.}c_2\} \mid
	\{\mathtt{v2.}c_1,\mathtt{v2.}c_2\} \in \mathtt{a0} \}$ \\
\> \> \texttt{share1 = }
    $\{ \{\mathtt{v1.}c_1,v.c_2\} \mid
	\{\mathtt{v2.}c_1,v.c_2\} \in \mathtt{a0} \}$ \\
\> \texttt{in} \\
\> \> $ \mathtt{a0} \cup \mathtt{self1} \cup \mathtt{share1}  $ \\
\texttt{alias (DerefEq v1 v2) a0 = }\>\>\>\>\>\>\>\>\>\>\>\>\>\texttt{-- *v1 = v2}\\
\> \texttt{let} \\
\> \> \texttt{self1 = }
    $\{\{\mathtt{v1.[Ref.1]},\mathtt{v1.[Ref.1]}\}\} ~\cup$ \\
\>\>\>\>\>
    $\{ \{\mathtt{fc (v1.(Ref.1:}c_1\mathtt{))},
		\mathtt{fc (v1.(Ref.1:}c_2\mathtt{))}\} \mid
	\{\mathtt{v2.}c_1,\mathtt{v2.}c_2\} \in \mathtt{a0} \}$ \\
\> \> \texttt{share1 = }
    $\{ \{\mathtt{fc (v1.(Ref.1:}c_1\mathtt{))},v.c_2\} \mid
	\{\mathtt{v2.}c_1,v.c_2\} \in \mathtt{a0} \}$ \\
\> \texttt{in} \\
\> \> $ \mathtt{a0} \cup \mathtt{self1} \cup \mathtt{share1}  $ \\
\end{tabbing}

Sequencing is handled by function composition.  To bind a fresh variable
\texttt{v1} to a variable \texttt{v2} the self-aliasing of \texttt{v2}
(including aliasing between different components of \texttt{v2})
is duplicated for \texttt{v1} and the aliasing for each component
of \texttt{v2} (which includes self-aliasing) is duplicated for \texttt{v1}.
Binding \verb@*v1@ to
\texttt{v2} is done in a similar way, but the components of \texttt{v1}
must have \texttt{Ref.1} prepended to them and the result folded, and
the \verb@[Ref.1]@ component of \texttt{v1} self-aliases.
Folding is
only needed for the rare case of types with recursion through \texttt{Ref}.

\pagebreak[4]
\begin{tabbing}
mm\=mm\=mm\=mm\=mm\=mm\=mm\=mm\=mm\=mm\=mm\=mm\=mm\=mm\kill
\texttt{alias (Assign v1 v2) a0 = }\>\>\>\>\>\>\>\>\>\>\>\>\>\texttt{-- *v1 := v2 }\\
\> \texttt{let} \\
\> \> \texttt{-- al = possible aliases for v1.[Ref.1] }\\
\> \> \texttt{al = }
    $\{ v_a.c_a \mid
	\{\mathtt{v1.[Ref.1]},v_a.c_a\} \in \mathtt{a0} \}$ \\
\> \> \texttt{-- (live variables in al, which includes v1, must be }\\
\> \> \texttt{-- annotated with ! and must not share with abstract) }\\
\> \> \texttt{self1al = }
    $\{ \{\mathtt{fc (}v_a.(c_a\mathtt{++}c_1)\mathtt{)},
		\mathtt{fc (}v_b.(c_b\mathtt{++}c_2)\mathtt{)}\} \mid$\\
\>\>\>\>\>\>\>$v_a.c_a \in \texttt{al} \wedge v_b.c_b \in \texttt{al} \wedge
	\{\mathtt{v2.}c_1,\mathtt{v2.}c_2\} \in \mathtt{a0} \}$ \\
\> \> \texttt{share1al = }
    $\{ \{\mathtt{fc (}v_a.(c_a\mathtt{++}c_1)\mathtt{)},v.c_2\} \mid$ \\
\>\>\>\>\>\>\>$v_a.c_a \in \texttt{al} \wedge
	\{\mathtt{v2.}c_1,v.c_2\} \in \mathtt{a0} \}$\\
\> \texttt{in if v1 is a mutable parameter then}\\
\>\>\> $ \mathtt{a0} \cup \mathtt{self1al} \cup \mathtt{share1al} $ \\
\> \> \texttt{else let }\\
\>\> \> \texttt{-- old1 = old aliases for v1, which can be removed }\\
\>\> \> \texttt{old1 = }
    $\{ \{\mathtt{v1.(Ref.1:}d:c_1),v.c_2\} \mid
	\{\mathtt{v1.(Ref.1:}d:c_1),v.c_2\} \in \mathtt{a0}\}$ \\
\>\> \texttt{in }$ (\mathtt{a0} \setminus \texttt{old1}) \cup
\mathtt{self1al} \cup \mathtt{share1al} $ \\
\end{tabbing}

Assignment to an existing variable differs from binding a fresh variable
in three ways.  First, self-sharing for \verb@v1.[Ref.1]@ is not
added since it already exists.  Second, \verb@v1.[Ref.1]@ may
alias several variable components (the live subset of these variables
must be annotated with ``!'' on the
assignment statement; checking such annotations is a primary purpose
of the analysis).  All these variables end up sharing with \texttt{v2}
and what \texttt{v2} shares with (via \texttt{share1al}) plus themselves
and each other (via \texttt{self1al}).  The components must be concatenated
and folded appropriately.  Third, if \texttt{v1} is not a mutable
parameter the existing sharing with a path strictly longer than
\texttt{[Ref.1]} (that is, paths of the form $\mathtt{Ref.1:}d:c_1$)
can safely be removed, improving precision.
The component \texttt{v1.[Ref.1]} represents the single memory word
that is overwritten and whatever the old contents shared with is no
longer needed to describe the sharing for \texttt{v1}.  For mutable
parameters the old value may share with variables from the calling
context and we retain this information, as explained later.
Consider the example below, where \texttt{t} and \texttt{ts} are as
before and local variables \texttt{v1} and \texttt{v3} are references to
the element of \texttt{ts}. The value assigned, \texttt{v2}, is
\texttt{RNode 3 (Cons (RNode 4 Nil) Nil)}.

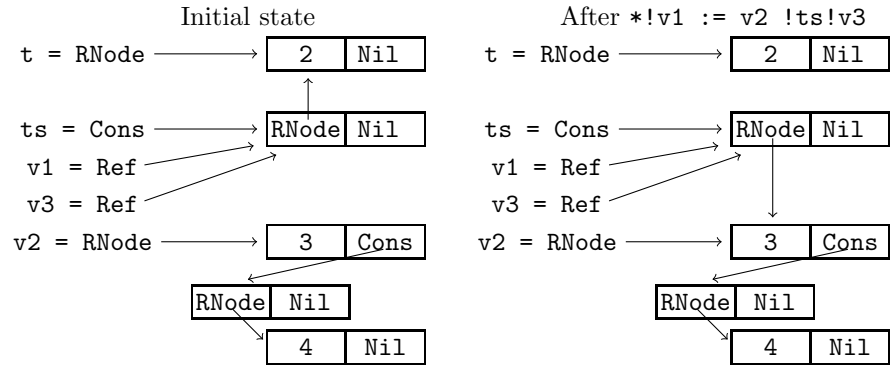
\begin{minipage}[c]{0.5\textwidth}
\noindent
\hspace*{-2em}\
\begin{tikzpicture}
\begin{scope}[scale=1.0]
\node at (2.2,4.2) (l1) {Initial state};
\node at (0,3.7) (c1) {\texttt{t = RNode}};
\node at (3.5,3.7) (c2)
        {\framebox[3em]{\texttt{ 2   }}\framebox[3em]{\texttt{Nil  }}};
\node at (3.0,3.5) (c2a) {};
\draw[->] (c1) -- (c2);
\node at (0,2.7) (d1) {\texttt{ts = Cons}};
\node at (3.5,2.7) (d2)
        {\framebox[3em]{\texttt{RNode}}\framebox[3em]{\texttt{Nil  }}};
\node at (3.0,2.7) (d2a) {};
\draw[->] (d1) -- (d2);
\node at (0,2.2) (e1) {\texttt{v1 = Ref}};
\node at (0.7,2.2) (e1a) {};
\node at (2.4,2.5) (e1b) {};
\draw[->] (e1a) -- (e1b);
\node at (0,1.7) (g1) {\texttt{v3 = Ref}};
\node at (0.7,1.7) (g1a) {};
\node at (2.7,2.4) (g1b) {};
\draw[->] (g1a) -- (g1b);
\node at (0,1.2) (f1) {\texttt{v2 = RNode}};
\node at (3.5,1.2) (f2)
        {\framebox[3em]{\texttt{ 3   }}\framebox[3em]{\texttt{Cons}}};
\node at (2.8,1.2) (f2a) {};
\draw[->] (f1) -- (f2);
\node at (3.0,1.4) (f2c) {};
\node at (2.5,0.4) (h2)
        {\framebox[3em]{\texttt{RNode}}\framebox[3em]{\texttt{Nil}}};
\node at (3.5,-0.2) (h3)
        {\framebox[3em]{\texttt{4}}\framebox[3em]{\texttt{Nil}}};
\draw[->] (4.0,1.1) -- (2.2,0.7);
\draw[->] (2.0,0.3) -- (2.4,-0.1);
\draw[->] (d2a) -- (c2a);
\end{scope}
\end{tikzpicture}
\end{minipage}
\begin{minipage}[c]{0.5\textwidth}
\noindent
\hspace*{-2em}\
\begin{tikzpicture}
\begin{scope}[scale=1.0]
\node at (2.2,4.2) (l1) {After \texttt{*!v1 := v2 !ts!v3}};
\node at (0,3.7) (c1) {\texttt{t = RNode}};
\node at (3.5,3.7) (c2)
        {\framebox[3em]{\texttt{ 2   }}\framebox[3em]{\texttt{Nil  }}};
\node at (3.0,3.5) (c2a) {};
\draw[->] (c1) -- (c2);
\node at (0,2.7) (d1) {\texttt{ts = Cons}};
\node at (3.5,2.7) (d2)
        {\framebox[3em]{\texttt{RNode}}\framebox[3em]{\texttt{Nil  }}};
\node at (3.0,2.7) (d2a) {};
\draw[->] (d1) -- (d2);
\node at (0,2.2) (e1) {\texttt{v1 = Ref}};
\node at (0.7,2.2) (e1a) {};
\node at (2.4,2.5) (e1b) {};
\draw[->] (e1a) -- (e1b);
\node at (0,1.7) (g1) {\texttt{v3 = Ref}};
\node at (0.7,1.7) (g1a) {};
\node at (2.7,2.4) (g1b) {};
\draw[->] (g1a) -- (g1b);
\node at (0,1.2) (f1) {\texttt{v2 = RNode}};
\node at (3.5,1.2) (f2)
        {\framebox[3em]{\texttt{ 3   }}\framebox[3em]{\texttt{Cons}}};
\node at (2.8,1.2) (f2a) {};
\draw[->] (f1) -- (f2);
\node at (3.0,1.4) (f2c) {};
\node at (2.5,0.4) (h2)
        {\framebox[3em]{\texttt{RNode}}\framebox[3em]{\texttt{Nil}}};
\node at (3.5,-0.2) (h3)
        {\framebox[3em]{\texttt{4}}\framebox[3em]{\texttt{Nil}}};
\draw[->] (4.0,1.1) -- (2.2,0.7);
\draw[->] (2.0,0.3) -- (2.4,-0.1);
\draw[->] (d2a) -- (f2c);
\end{scope}
\end{tikzpicture}
\end{minipage}

There is aliasing of \verb@v1.[Ref.1]@, \verb@v3.[Ref.1]@ and
\verb@ts.[Cons.1]@ so all these variables have the sharing of \texttt{v2}
and self-sharing added.  Generally we must also add sharing between all
pairs of these variables.  For example,
\verb@{ts.[Cons.1],@
\verb@v3.[Ref.1,RNode.2,Cons.1]}@
must be added because the \texttt{Cons} component of \texttt{v3} did
not previously exist.  The old sharing of \texttt{v1} with \texttt{t} is
discarded.  Note that we cannot discard the old sharing of \texttt{ts} and
\texttt{v3} with \texttt{t} for two reasons.  First, no \emph{definite}
aliasing information is maintained, so we cannot be sure \texttt{v3}
or \texttt{ts} are modified at all.  Second, the assignment updates
only one memory word whereas there may be other words also represented
by \texttt{ts.[Cons.1]}.  In some cases the old sharing of \texttt{v1}
is discarded and immediately added again.  Consider the following example,
which creates a cyclic list.

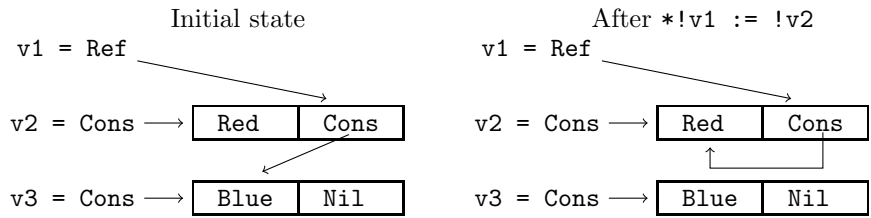
\begin{minipage}[c]{0.5\textwidth}
\noindent
\hspace*{-2em}\
\begin{tikzpicture}
\begin{scope}[scale=1.0]
\node at (2.2,2.9) (l1) {Initial state};
\node at (0,2.5) (v1) {\texttt{v1 = Ref}};
\node at (0,1.5) (v2) {\texttt{v2 = Cons}};
\node at (3.0,1.5) (c2)
        {\framebox[4em]{\texttt{Red }}\framebox[4em]{\texttt{Cons}}};
\draw[->] (v2) -- (c2);
\node at (3.5,1.8) (tt) {};
\draw[->] (v1) -- (tt);
\node at (3.8,1.5) (t) {};
\node at (2.3,1.3) (h) {};
\node at (0,0.5) (v3) {\texttt{v3 = Cons}};
\node at (3.0,0.5) (c3)
        {\framebox[4em]{\texttt{Blue}}\framebox[4em]{\texttt{Nil }}};
\draw[->] (v3) -- (c3);
\node at (2.4,0.8) (h3) {};
\node at (3.8,1.4) (t2) {};
\draw[->] (t2) -- (h3);
\node at (2.0,0.6) (bot) { };
\end{scope}
\end{tikzpicture}
\end{minipage}
\begin{minipage}[c]{0.5\textwidth}
\noindent
\hspace*{-2em}\
\begin{tikzpicture}
\begin{scope}[scale=1.0]
\node at (2.2,2.9) (l1) {After \texttt{*!v1 := !v2}};
\node at (0,2.5) (v1) {\texttt{v1 = Ref}};
\node at (0,1.5) (v2) {\texttt{v2 = Cons}};
\node at (3.0,1.5) (c2)
        {\framebox[4em]{\texttt{Red }}\framebox[4em]{\texttt{Cons}}};
\draw[->] (v2) -- (c2);
\node at (3.5,1.8) (tt) {};
\draw[->] (v1) -- (tt);
\node at (3.8,1.5) (t) {};
\node at (2.3,1.3) (h) {};
\draw[->] (t) -- (3.8,0.9) -- (2.3,0.9) -- (h);
\node at (0,0.5) (v3) {\texttt{v3 = Cons}};
\node at (3.0,0.5) (c3)
        {\framebox[4em]{\texttt{Blue}}\framebox[4em]{\texttt{Nil }}};
\draw[->] (v3) -- (c3);
\node at (2.4,0.8) (h3) {};
\node at (3.8,1.4) (t2) {};
\node at (2.0,0.6) (bot) { };
\end{scope}
\end{tikzpicture}
\end{minipage}

The sharing between \texttt{v1} and \texttt{v3} is discarded but
added again (via \texttt{share1al}) because \texttt{v2} also shares with
\texttt{v3}. Correctness of the algorithm when cyclic terms are created
depends on the abstract domain we use.  A more expressive domain could
distinguish between different cons cells in a list.  For example, if
types are ``folded'' at the third level of recursion rather than the
first, the domain can distinguish three classes of cons cells, where the
distance from the first cons cell, modulo three, is zero, one or two.
For a cyclic list with a single cons cell, that cons cell must be in all
three classes and our algorithm would need modification to achieve this.
However, in our domain types are folded at the first level of recursion so
we have a unique folded path for each memory cell in cyclic data structure
(cyclic terms can only be created with recursive types).  There is no
distinction between the first and second cons cell in a list, for example.

\begin{tabbing}
mm\=mm\=mm\=mm\=mm\=mm\=mm\=mm\=mm\=mm\=mm\=mm\=mm\=mm\kill
\texttt{alias (DC v dc [}$v_1,\ldots v_N$\texttt{]) a0 =
}\>\>\>\>\>\>\>\>\>\>\>\>\>\texttt{-- v = Dc v1...vN}\\
\> \texttt{let} \\
\> \> \texttt{self1 = }
    $ \{\{\mathtt{fc (v.[dc.}i]),\mathtt{fc (v.[dc.}i])\}
          \mid 1 \leq i \leq N \} ~~\cup $ \\
\>\>\>\>\>\>$ \{ \{\mathtt{fc (v.(dc.}i\mathtt{:}c_1)),
	\mathtt{fc (v.(dc.}j\mathtt{:}c_2)) \} \mid
	\{v_i.c_1,v_j.c_2\} \in \mathtt{a0} \}$ \\
\> \> \texttt{share1 = }
    $ \{ \{\mathtt{fc (v.(dc.}i\mathtt{:}c_1)),
	w.c_2 \} \mid \{v_i.c_1,w.c_2\} \in \mathtt{a0} \} $ \\
\> \texttt{in} \\
\> \> $ \mathtt{a0} \cup \mathtt{self1} \cup \mathtt{share1}  $ \\
\end{tabbing}

The \texttt{DerefEq} case can be seen as equivalent to \texttt{v1 =
Ref v2} and binding a variable to a data constructor with $N$ variable
arguments is a generalisation.  If there are multiple $v_i$ that
share, the corresponding components of \texttt{v} must also share; these
pairs are included in \texttt{self1}. 

\pagebreak[4]
\begin{tabbing}
mm\=mm\=mm\=mm\=mm\=mm\=mm\=mm\=mm\=mm\=mm\=mm\=mm\=mm\kill
\texttt{alias (EqDeref v1 v2) a0 = }\>\>\>\>\>\>\>\>\>\>\>\>\>\texttt{-- v1 = *v2}\\
\> \texttt{let} \\
\> \> \texttt{self1 = }
    $\{ \{\mathtt{v1.}c_1,\mathtt{v1.}c_2\} \mid
	\{\mathtt{fc(v2.(Ref.1:}c_1\mathtt{))},\mathtt{fc(v2.(Ref.1:}c_2\mathtt{))}\} \in \mathtt{a0} \}$ \\
\> \> \texttt{share1 = }
    $\{ \{\mathtt{v1.}c_1,v.c_2\} \mid
	\{\mathtt{fc(v2.(Ref.1:}c_1\mathtt{))},v.c_2\} \in \mathtt{a0} \}$ \\
\> \> \texttt{empty1 = }
    $\{ \{$\texttt{v1.[]}$,v.c\} \mid
	\{$\texttt{v1.[]}$,v.c\} \in (\mathtt{self1} \cup \mathtt{share1}) \}$ \\
\> \texttt{in} \\
\> \> \texttt{if the type of v1 has a [] component then }\\
\>\> \> $ \mathtt{a0} \cup \mathtt{self1} \cup \mathtt{share1}  $ \\
\> \> \texttt{else     --- avoid bogus sharing with empty component}\\
\>\> \> $ (\mathtt{a0} \cup \mathtt{self1} \cup \mathtt{share1})
\setminus \mathtt{empty1}  $ \\
\end{tabbing}

The \texttt{EqDeref} case is similar to the inverse of \texttt{DerefEq} in
that we are removing \texttt{Ref.1} rather than prepending it (the
definition implicitly uses the inverse of \texttt{fc}).  However,
if the empty component results we must check that such a component exists
for the type of \texttt{v1}.

\begin{tabbing}
mm\=mm\=mm\=mm\=mm\=mm\=mm\=mm\=mm\=mm\=mm\=mm\=mm\=mm\kill
\texttt{alias (App v f [}$v_1,\ldots v_N$\texttt{]) a0 =
}\>\>\>\>\>\>\>\>\>\>\>\>\>\texttt{-- v = f v1...vN}\\
\> \texttt{let} \\
\> \> ``\texttt{f(}$w_1, \ldots w_{K+N}) = r$'' \texttt{is used to declare
sharing for f}\\
\> \> \texttt{mut = the arguments that are declared mutable } \\
\> \> \texttt{post = the postcondition of f along with the sharing for } \\
\>\>\>\>\>  \texttt{mutable arguments from the precondition, }\\
\>\>\>\>\>  \texttt{with parameters and result renamed with}\\
\>\>\>\>\>  \texttt{f}$.[\texttt{Cl}.K],\ldots
	\texttt{f}.[\texttt{Cl}.1],v_1,\ldots v_N$ \texttt{and v,
respectively} \\
\> \> \texttt{-- (the renamed precondition of f must be a subset of a0,} \\
\> \> \texttt{-- and mutable arguments of f and live variables they
share} \\
\> \> \texttt{-- with must be annotated with ! and must not share with}\\
\> \> \texttt{-- abstract)}\\
\> \> \texttt{-- selfc+sharec needed for possible closure creation} \\
\> \> \texttt{selfc = }
	$\{\{\mathtt{v.[Cl.}i\mathtt{]},\mathtt{v.[Cl.}i\mathtt{]}\} \mid 1\leq i\leq N
\} ~\cup$\\
\>\>\>\>\>
	$\{\{\mathtt{v.((Cl.}(N+1-i)\mathtt{):}c_1),\mathtt{v.((Cl.}(N+1-j)\mathtt{):}c_2)\} ~\mid
$\\ \>\>\>\>\>\>$
	\{v_i.c_1,v_j.c_2\} \in
\mathtt{a0} \}~\cup$ \\
\>\>\>\>\>
    $\{
\{\mathtt{v.((Cl.(}i+N\mathtt{)):}c_1),\mathtt{v.((Cl.(}j+N\mathtt{)):}c_2)\} \mid
$\\ \>\>\>\>\>\>$
	\{\mathtt{f.((Cl.}i\mathtt{):}c_1),\mathtt{f.((Cl.}j\mathtt{):}c_2)\} \in \mathtt{a0} \}$ \\
\> \> \texttt{sharec = }
	$\{\{\mathtt{v.((Cl.}(N+1-i)\mathtt{):}c_1),x.c_2\} ~\mid
	\{v_i.c_1,x.c_2)\} \in
\mathtt{a0} \}~\cup$ \\
\>\>\>\>\>
    $\{ \{\mathtt{v.((Cl.(}i+N\mathtt{)):}c_1),x.c_2\} \mid
	\{\mathtt{f.((Cl.}i\mathtt{):}c_1),x.c_2\} \in \mathtt{a0} \}$ \\
\> \> \texttt{-- postt+postm needed for possible function call} \\
\> \> \texttt{postt = }
    $\{ \{x_1.c_1,x_3.c_3\} \mid
	\{x_1.c_1,x_2.c_2\} \in \mathtt{post} \wedge
	\{x_2.c_2,x_3.c_3\} \in \mathtt{a0} \}$ \\
\> \> \texttt{postm = }
    $\{ \{x_1.c_1,x_2.c_2\} \mid
	\{x_1.c_1,v_i.c_3\} \in \mathtt{a0} \wedge
	\{x_2.c_2,v_j.c_4\} \in \mathtt{a0}~ \wedge$\\
\>\>\>\>\>	$
	\{v_i.c_3,v_j.c_4\} \in \mathtt{post} \wedge
	v_i \in \mathtt{mut} \wedge v_j \in \mathtt{mut} \}$ \\
\> \texttt{in} \\
\> \>    $\mathtt{a0} \cup \mathtt{selfc} \cup \mathtt{sharec} \cup \mathtt{postt} \cup \mathtt{postm} $ \\
\end{tabbing}

For many \texttt{App} occurrences the function is known statically and we can
determine if the function is actually called or a closure is created
instead.  However, in general we must assume either could happen and
add sharing for both.  If a closure is created, the first $N$ closure
arguments share with the $N$ arguments of the function call and any
closure arguments of \texttt{f} share with additional closure arguments of
the result (this requires renumbering of these arguments).

Analysis of
function calls relies on the sharing and mutability information attached
to all arrow types.  Because Pawns uses the syntax of statements to
express pre- and post-conditions, our implementation uses the sharing
analysis algorithm to derive an explicit alias set representation
(currently this is done recursively, with the level of recursion limited
by the fact than pre- and post-conditions must not contain function
calls).  Here we ignore the details of how the alias set representation
is obtained.  The compiler also uses the sharing information immediately
before an application to check that the precondition is satisfied,
all required ``!'' annotations are present and abstract variables are
not modified.

Given that the precondition is satisfied, the execution of a function
results in sharing of parameters that is a subset of the union of
the declared pre- and post-conditions (we assume the induction hypothesis
holds for the sub-computation, which has a smaller depth of recursion).
However, any sharing between non-mutable arguments that exists
immediately after the call must exist before the call.
The analysis algorithm does not add sharing between non-mutable
arguments in the precondition as doing so would unnecessarily restrict
how ``high level'' and ``low level'' code can be mixed.  It is important
we can pass a variable to a function that allows an abstract argument
without the analysis concluding the variable subsequently shares with
\texttt{abstract}, and therefore cannot be updated.  Thus
\texttt{post} is just the declared
postcondition plus the subset of the precondition which involves mutable
parameters of the function, renamed appropriately.
The last $N$ formal parameters,  
$w_{K+1} \ldots w_{K+N}$ are renamed as the arguments of the call,
$v_1 \ldots v_N$ and the formal result $r$ is renamed \texttt{v}.
The formal parameters $w_{1} \ldots w_{K}$ represent closure arguments
$K \ldots 1$ of \texttt{f}.  Thus a variable component such as
$w_{1}$\texttt{.[Cons.1]} is renamed \texttt{f.[Cl.}$K$\texttt{,Cons.1]}.

It is also necessary to include one step of transitivity in the sharing
information: if variable components $x_1.c_1$ and $x_2.c_2$ alias in
\texttt{post} and $x_2.c_2$ and $x_3.c_3$ (may) alias before the function
call, we add an alias of $x_1.c_1$ and $x_3.c_3$ (in \texttt{postt}).
Function parameters are proxies for
the argument variables as well as any variable components they may alias
and when functions are analysed these aliases are not known.
This is why the transitivity step is needed, and why mutable parameters
also require special treatment.  If before the call, $x_1.c_1$ and $x_2.c_2$
may alias with mutable parameter components $v_i.c_3$ and $v_j.c_4$,
respectively, and the two mutable parameter components alias in
\texttt{post} then $x_1.c_1$ and $x_2.c_2$ may alias after the call;
this is added in \texttt{postm}.  Consider the example below, where
we have a pair \texttt{v1} (of references to references to integers)
and variables \texttt{x} and \texttt{y} share with the two elements
of \texttt{v1}, respectively.  When \texttt{v1} is passed to function
\texttt{f1}
as a mutable parameter, sharing between \texttt{x} and \texttt{y} is
introduced.  The sharing of the mutable parameter in the postcondition,
\texttt{\{v1.[Pair.1,Ref.1,Ref.1], v1.[Pair.2,Ref.1,Ref.1]\}}, results
in sharing between \texttt{x} and \texttt{y} being added in the analysis.

\noindent
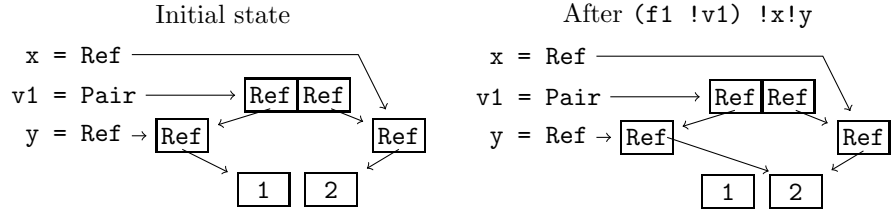
\begin{minipage}[c]{0.5\textwidth}
\begin{tikzpicture}
\begin{scope}[scale=0.9]
\node at (2.2,3.6) (l1) {Initial state};
\node at (0,3.0) (x) {\texttt{x = Ref}};
\node at (2.8,1.0) (ba)
        {\framebox[2em]{\texttt{1}}};
\node at (3.8,1.0) (bb)
        {\framebox[2em]{\texttt{2}}};
\node at (0,2.4) (v1) {\texttt{v1 = Pair}};
\node at (3.3,2.4) (v1a) {\framebox[2em]{\texttt{Ref}}\framebox[2em]{\texttt{Ref}}};
\draw[->] (v1) -- (v1a);
\node at (0,1.8) (y) {\texttt{y = Ref}};
\node at (1.6,1.8) (ya) {\framebox[2em]{\texttt{Ref}}};
\node at (4.8,1.8) (yb) {\framebox[2em]{\texttt{Ref}}};
\draw[->] (y) -- (ya);
\draw[->] (1.6,1.6) -- (ba);
\draw[->] (4.8,1.6) -- (bb);
\draw[->] (2.9,2.2) -- (ya);
\draw[->] (3.8,2.2) -- (yb);
\draw[->] (x) -- (4.2,3.0) -- (yb);
\node at (0,0.1) (bot) {};
\end{scope}
\end{tikzpicture}
\end{minipage}
\begin{minipage}[c]{0.5\textwidth}
\begin{tikzpicture}
\begin{scope}[scale=0.9]
\node at (2.2,3.6) (l1) {After \texttt{(f1 !v1) !x!y}};
\node at (0,3.0) (x) {\texttt{x = Ref}};
\node at (2.8,1.0) (ba)
        {\framebox[2em]{\texttt{1}}};
\node at (3.8,1.0) (bb)
        {\framebox[2em]{\texttt{2}}};
\node at (0,2.4) (v1) {\texttt{v1 = Pair}};
\node at (3.3,2.4) (v1a) {\framebox[2em]{\texttt{Ref}}\framebox[2em]{\texttt{Ref}}};
\draw[->] (v1) -- (v1a);
\node at (0,1.8) (y) {\texttt{y = Ref}};
\node at (1.6,1.8) (ya) {\framebox[2em]{\texttt{Ref}}};
\node at (4.8,1.8) (yb) {\framebox[2em]{\texttt{Ref}}};
\draw[->] (y) -- (ya);
\draw[->] (1.9,1.8) -- (3.4,1.3);
\draw[->] (4.8,1.6) -- (bb);
\draw[->] (2.9,2.2) -- (ya);
\draw[->] (3.8,2.2) -- (yb);
\draw[->] (x) -- (4.2,3.0) -- (yb);
\node at (0,0.1) (bot) {};
\end{scope}
\end{tikzpicture}
\end{minipage}

\begin{verbatim}
f1 :: Pair (Ref (Ref Int)) -> ()
    sharing f1 !v1 = _
    pre nosharing
    post *a = *b; v1 = Pair a b
f1 !v1 =
    case v1 of (Pair rr1 rr2) -> *rr1 := *rr2 !v1
\end{verbatim}

The need to be conservative with the sharing of mutable parameters
in the analysis of function definitions (the special treatment in
\texttt{Assign}) is illustrated by the example below.  Consider the
initial state, with variables \texttt{v1} and \texttt{v2} which share
with \texttt{x} and \texttt{y}, respectively.  After \texttt{f2} is
called \texttt{x} and \texttt{y} share, even though the parameters
\texttt{v1} and \texttt{v2} do not share at any point in the execution
of \texttt{f2}.  If mutable parameters were not treated specially in
the \texttt{Assign} case, \texttt{nosharing} would be accepted as the
postcondition of \texttt{f2} and the analysis of the call to \texttt{f2}
would then be incorrect.  The sharing is introduced between memory cells
that were once shared with \texttt{v1} and others that were once shared
with \texttt{v2}.  Thus in our algorithm, the sharing
of mutable parameters reflects all memory cells that are reachable from
the parameters during the execution of the function.  Where the mutable
parameters are assigned in \texttt{f2}, the sharing of the parameters'
previous values (\texttt{rr1} and \texttt{rr2}) is retained.
Thus when the final assignment is processed, sharing between the
parameters is added and this must be included in the postcondition.
Although this assignment does not modify \texttt{v1} or \texttt{v2}, the
``!'' annotations are necessary and alert the reader to potential
modification of variables that shared with the parameters when the
function was called.

\noindent
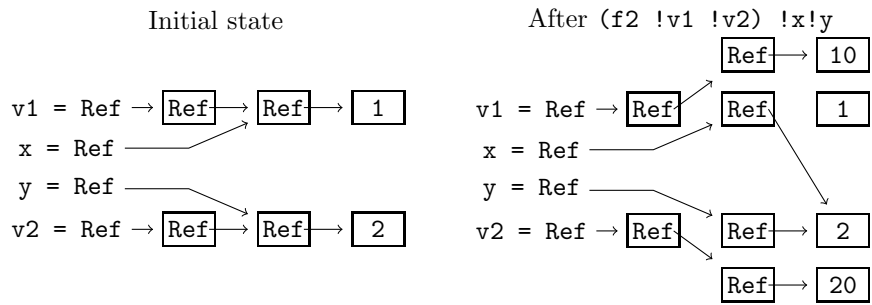
\begin{minipage}[c]{0.5\textwidth}
\begin{tikzpicture}
\begin{scope}[scale=0.9]
\node at (2.2,4.3) (l1) {Initial state};
\node at (0,3.0) (v1) {\texttt{v1 = Ref}};
\node at (1.8,3.0) (v1a)
        {\framebox[2em]{\texttt{Ref}}};
\node at (3.2,3.0) (v1b)
        {\framebox[2em]{\texttt{Ref}}};
\node at (4.6,3.0) (v1c)
        {\framebox[2em]{\texttt{1}}};
\draw[->] (v1) -- (v1a);
\draw[->] (3.5,3.0) -- (v1c);
\draw[->] (2.1,3.0) -- (v1b);
\node at (0,1.2) (v2) {\texttt{v2 = Ref}};
\node at (1.8,1.2) (v2a)
        {\framebox[2em]{\texttt{Ref}}};
\node at (3.2,1.2) (v2b)
        {\framebox[2em]{\texttt{Ref}}};
\node at (4.6,1.2) (v2c)
        {\framebox[2em]{\texttt{2}}};
\draw[->] (v2) -- (v2a);
\draw[->] (3.5,1.2) -- (v2c);
\draw[->] (2.1,1.2) -- (v2b);
\node at (0,2.4) (x) {\texttt{x = Ref}};
\draw[->] (x) -- (1.8,2.4) -- (v1b);
\node at (0,1.8) (y) {\texttt{y = Ref}};
\draw[->] (y) -- (1.8,1.8) -- (v2b);
\node at (0,0.1) (bot) {};
\end{scope}
\end{tikzpicture}
\end{minipage}
\begin{minipage}[c]{0.5\textwidth}
\noindent
\begin{tikzpicture}
\begin{scope}[scale=0.9]
\node at (2.2,4.3) (l1) {After \texttt{(f2 !v1 !v2) !x!y}};
\node at (0,3.0) (v1) {\texttt{v1 = Ref}};
\node at (1.8,3.0) (v1a)
        {\framebox[2em]{\texttt{Ref}}};
\node at (3.2,3.0) (v1b)
        {\framebox[2em]{\texttt{Ref}}};
\node at (4.6,3.0) (v1c)
        {\framebox[2em]{\texttt{1}}};
\draw[->] (v1) -- (v1a);
\draw[->] (3.5,3.0) -- (v2c);
\node at (3.2,3.8) (n1b)
        {\framebox[2em]{\texttt{Ref}}};
\node at (4.6,3.8) (n1c)
        {\framebox[2em]{\texttt{10}}};
\draw[->] (2.1,3.0) -- (n1b);
\draw[->] (3.5,3.8) -- (n1c);
\node at (0,1.2) (v2) {\texttt{v2 = Ref}};
\node at (1.8,1.2) (v2a)
        {\framebox[2em]{\texttt{Ref}}};
\node at (3.2,1.2) (v2b)
        {\framebox[2em]{\texttt{Ref}}};
\node at (4.6,1.2) (v2c)
        {\framebox[2em]{\texttt{2}}};
\draw[->] (v2) -- (v2a);
\draw[->] (3.5,1.2) -- (v2c);
\node at (3.2,0.4) (n2b)
        {\framebox[2em]{\texttt{Ref}}};
\node at (4.6,0.4) (n2c)
        {\framebox[2em]{\texttt{20}}};
\draw[->] (2.1,1.2) -- (n2b);
\draw[->] (3.5,0.4) -- (n2c);
\node at (0,2.4) (x) {\texttt{x = Ref}};
\draw[->] (x) -- (1.8,2.4) -- (v1b);
\node at (0,1.8) (y) {\texttt{y = Ref}};
\draw[->] (y) -- (1.8,1.8) -- (v2b);
\node at (0,0.1) (bot) {};
\end{scope}
\end{tikzpicture}
\end{minipage}

\pagebreak[4]
\begin{verbatim}
f2 :: Ref (Ref (Ref Int)) -> Ref (Ref (Ref Int)) -> ()
    sharing f2 !v1 !v2 = _
    pre nosharing
    post **v1 = **v2
f2 !v1 !v2 =
    *r10 = 10           -- ref to new cell containing 10
    *rr10 = r10         -- ref to above ref
    *r20 = 20           -- ref to new cell containing 20
    *rr20 = r20         -- ref to above ref
    rr1 = *v1           -- save *v1
    rr2 = *v2           -- save *v2
    *!v1 := rr10        -- update *v1 with Ref (Ref 10)
    *!v2 := rr20        -- update *v2 with Ref (Ref 20)
    *rr1 := *rr2 !v1!v2 -- can create sharing at call
\end{verbatim}

\begin{tabbing}
mm\=mm\=mm\=mm\=mm\=mm\=mm\=mm\=mm\=mm\=mm\=mm\=mm\=mm\kill
\texttt{alias Error a0 = } $\emptyset$\>\>\>\>\>\>\>\>\>\>\>\>\>\texttt{-- error}\\
\texttt{alias (Case v [}$(p_1,s_1),\ldots(p_N,s_N$\texttt{)]) a0 = }
	\>\>\>\>\>\>\>\>\>\>\>\>\>\texttt{-- case v of ...}\\
\> \texttt{let} \\
\> \> \texttt{old = }
    $\{ \{\mathtt{v.}c_1,v_2.c_2\} \mid
	\{\mathtt{v.}c_1,v_2.c_2\} \in \mathtt{a0} \}$ \\
\> \texttt{in} \\
\> \>    $
\bigcup_{1\leq i\leq
N} \mathtt{aliasCase \ a0 \ old \ v}~ p_i~s_i $ \\
\end{tabbing}
\begin{tabbing}
mm\=mm\=mm\=mm\=mm\=mm\=mm\=mm\=mm\=mm\=mm\=mm\=mm\=mm\kill
\texttt{aliasCase a0 av v (Pat dc [}$v_1,\ldots v_N$\texttt{])\ s = }
	\>\>\>\>\>\>\>\>\>\>\>\>\>\texttt{-- (Dc *v1...*vN) -> s}\\
\> \texttt{let} \\
\> \> $\mathtt{avdc}$\texttt{\ =\ }
	$\{ \{\mathtt{fc(v.(dc.}i:c_1)),
	w.c_2\} \mid \{\mathtt{fc(v.(dc.}i:c_1)),w.c_2\} \in \mathtt{av}
\}$ \\
\> \> $\mathtt{rself}$\texttt{\ =\ }
    $\{\{v_i.\mathtt{[Ref.1]},v_i.\mathtt{[Ref.1]}\} \mid 1\leq i\leq N\}$ \\
\> \> $\mathtt{vishare}$\texttt{\ =\ }
    $\{ \{\mathtt{fc (}v_i.\mathtt{(Ref.1:}c_1)),
		\mathtt{fc (}v_j.\mathtt{(Ref.1:}c_2))\} \mid $\\
\>\>\>\>\>\> $\{\mathtt{fc(v.(dc.}i:c_1)),\mathtt{fc(v.(dc.}j:c_2))\}
\in \mathtt{av}
\}$ \\
\> \> $\mathtt{share}$\texttt{\ =\ }
    $\{ \{\mathtt{fc (}v_i.\mathtt{(Ref.1:}c_1)),
		w.c_2\} \mid
\{\mathtt{fc(v.(dc.}i:c_1)),w.c_2))\}
\in \mathtt{av}
\}$ \\
\> \texttt{in} \\
\> \> \texttt{alias s} $(
	\mathtt{rself} \cup \mathtt{vishare} \cup \mathtt{share}
	\cup
	(\texttt{a0} \setminus \texttt{av})
	\cup
	\texttt{avdc}
)$
\\
\end{tabbing}

For a case expression we return the union of the alias sets obtained for
each of the different branches.  For each branch we only keep sharing
information for the variable we are switching on that is compatible
with the data constructor in that branch (we remove all the old sharing,
\texttt{av}, and add the compatible sharing, \texttt{avdc}).
We implicitly use the inverse of \texttt{fc}.  To deal
with individual data constructors we consider pairs of components of
arguments $i$ and $j$ which may alias in order to compute possible
sharing between $v_i$ and $v_j$, including self-aliases when $i=j$.
The corresponding component of $v_i$ (prepended with \texttt{Ref} and
folded) may alias the component of $v_j$.  For example, if \texttt{v}
of type \texttt{RTrees} is matched with \texttt{Cons *v1 *v2} and
\verb@v.[]@ self-aliases, we need to find the components which fold
to \verb@v.[]@ (\verb@v.[Cons.2]@ and \verb@v.[Cons.1,RNode.2]@) in order
to compute the sharing for \texttt{v2} and \texttt{v1}.  Thus we compute
that \verb@v2.[Ref.1]@,
may alias \verb@v1.[Ref.1,RNode.2]@. This can occur if the
data structure is cyclic, such as the example below where \texttt{v} is
a list containing a single tree with 2 in the node and \texttt{v} as the
children (hence it represents a single infinite branch).  Note that
\verb@v1.[Ref.1,RNode.2]@ represents both the memory cell
containing the \texttt{Cons} pointer and the cell containing \texttt{Nil}.

\noindent
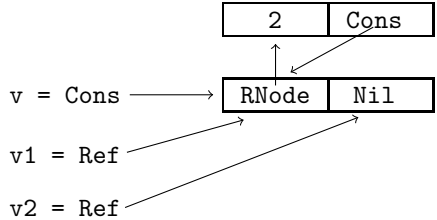
\begin{tikzpicture}
\begin{scope}[scale=1.0]
\node at (3.5,3.7) (c2)
        {\framebox[4em]{\texttt{ 2   }}\framebox[4em]{\texttt{Cons  }}};
\node at (2.8,3.5) (c2a) {};
\node at (0,2.7) (d1) {\texttt{v = Cons}};
\node at (3.5,2.7) (d2)
        {\framebox[4em]{\texttt{RNode}}\framebox[4em]{\texttt{Nil  }}};
\node at (2.8,2.7) (d2a) {};
\draw[->] (d1) -- (d2);
\node at (0,1.9) (e1) {\texttt{v1 = Ref}};
\node at (0.7,1.9) (e1a) {};
\node at (2.5,2.4) (e1b) {};
\draw[->] (e1a) -- (e1b);
\node at (0,1.2) (f1) {\texttt{v2 = Ref}};
\node at (2.8,1.2) (f2a) {};
\draw[->] (0.8,1.2) -- (3.9,2.4);
\draw[->] (4.1,3.6) -- (3.0,3.0);
\node at (2.8,1.4) (f2c) {};
\draw[->] (d2a) -- (c2a);
\end{scope}
\end{tikzpicture}

\begin{tabbing}
mm\=mm\=mm\=mm\=mm\=mm\=mm\=mm\=mm\=mm\=mm\=mm\=mm\=mm\kill
\texttt{alias (Instype v1 v2) a0 = }\>\>\>\>\>\>\>\>\>\>\>\>\>\texttt{-- v1 = v2::t}\\
\> \texttt{alias (EqVar v1 v2) a0} \\
\> \texttt{-- (if any sharing is introduced between v1 and v2,} \\
\> \texttt{-- v2 must not be indirectly updated later while live)}
\end{tabbing}

Type instantiation is dealt with in the same way as variable equality,
with the additional check that if any sharing is introduced, the variable
with the more general type is not implicitly updated later while still
live (it is sufficient to check there is no ``\texttt{!v2}'' annotation
attached to a later statement).

\section{Example}
\label{sec-example}

We now show how this sharing analysis algorithm is applied to the binary
search tree code given earlier.  We give a core Pawns version of each
function and the alias set before and after
each statement, plus an additional set at the end which is the union of
the pre- and post-conditions of the function.  To save space, we write
the alias set as a set of sets where each inner set represents all sets
containing exactly two of its members.  Thus $\{\{a,b,c\}\}$ represents a
set of six alias pairs: aliasing between all pairs of elements, including
self-aliases.  The return value is given by variable \texttt{ret}
and variables \texttt{absL} and \texttt{absT} are the versions of
\texttt{abstract} for type \texttt{Ints} and \texttt{Tree}, respectively.

\begin{verbatim}
list_bst xs =                       -- 0
    v1 = TNil                       -- 1
    *tp = v1                        -- 2
    list_bst_du xs !tp              -- 3
    ret = *tp                       -- 4
\end{verbatim}

We start with the precondition:
$a_0 = \{\{$\texttt{xs.[Cons.1], absL.[Cons.1]}$\}$,
$\{$\texttt{xs.[], absL.[]}$\}\}$.
Binding to a constant introduces no sharing so $a_1 = a_0$.
$a_2 = a_1$ $\cup$ $\{$\texttt{tp.[Ref.1]}$\}$.
The function call has precondition $a_0 \cup \{\{$\texttt{tp.[Ref.1]}$\},$
$\{$\texttt{tp.[Ref.1,Node.2]}$\}\}$, which is a superset of $a_2$.
Since \texttt{tp} is a mutable argument the
precondition sharing for \texttt{tp} is added:
$a_3 = a_2$ $\cup$ $\{\{$\texttt{tp.[Ref.1,} \texttt{Node.2]}$\}\}$.
The final sharing includes the return variable, \texttt{ret}:
$a_4 = a_3$ $\cup$ $\{\{$\texttt{ret.[],tp.[Ref.1]}$\}$,
$\{$\texttt{ret.[Node.2],tp.[Ref.1,Node.2]}$\}\}$.  After removing sharing
for the dead (local) variable \texttt{tp} we obtain
a subset of the union of the pre- and post-conditions,
which is
$a_0 \cup \{\{$\texttt{ret.[],absT.[]}$\},$
$\{$\texttt{ret.[Node.2]}, \texttt{absT.[Node.2]}$\}\}$.

\begin{verbatim}
list_bst_du xs !tp =                -- 0
    case xs of
    (Cons *v1 *v2) ->               -- 1
       x = *v1                      -- 2
       xs1 = *v2                    -- 3
       v3 = bst_insert_du x !tp     -- 4
       v4 = list_bst_du xs1 !tp     -- 5
       ret = v4                     -- 6
    Nil ->                          -- 7
       ret = ()                     -- 8
    -- after case                   -- 9
\end{verbatim}

We start with the precondition,
$a_0 = \{\{$\texttt{tp.[Ref.1]}$\}$, $\{$\texttt{tp.[Ref.1,Node.2]}$\}$,
$\{$\texttt{xs.[Cons.1],absL.[Cons.1]}$\},$
$\{$\texttt{xs.[],absL.[]}$\}\}$.
The \texttt{Cons} branch of the case introduces sharing for \texttt{v1} and
\texttt{v2}:
$a_1 = a_0$ $\cup$ $\{\{$\texttt{xs.[Cons.1]}, \texttt{absL.[Cons.1]},
\texttt{v1.[Ref.1]}, \texttt{v2.[Ref.1,Cons.1]}$\},$
$\{$\texttt{v2.[Ref.1],} \texttt{xs.[]}, \texttt{absL.[]}$\}\}$.
The list elements are atomic so $a_2 = a_1$.
The next binding makes the sharing of \texttt{xs1} and \texttt{xs} the
same:
$a_3 = a_2$ $\cup$ $\{\{$\texttt{v2.[Ref.1],} \texttt{xs.[]},
\texttt{xs1.[]},
\texttt{absL.[]}$\}$,
$\{$\texttt{v1.[Ref.1],} \texttt{xs.[Cons.1]}, \texttt{xs1.[Cons.1]}, \texttt{absL.[Cons.1]},
\texttt{v2.[Ref.1,Cons.1]}$\}\}$.
This can be simplified by removing the dead variables \texttt{v1} and
\texttt{v2}.
The precondition of the calls are satisfied and
$a_6 = a_5 = a_4 = a_3$.
For the \texttt{Nil} branch we remove the incompatible sharing for
\texttt{xs} from $a_0$:
$a_7 = \{\{$\texttt{tp.[Ref.1]}$\}$, $\{$\texttt{tp.[Ref.1,Node.2]}$\}$,
$\{$\texttt{absL.[Cons.1]}$\}$,
$\{$\texttt{absL.[]}$\}\}$
and $a_8 = a_7$.  Finally, $a_9 = a_6
\cup a_8$. This contains all the sharing for mutable parameter \texttt{tp} and,
ignoring local variables, is a
subset of the union of the pre- and post-conditions, $a_0$.

\begin{verbatim}
bst_insert_du x !tp =                          -- 0
    v1 = *tp                                   -- 1
    case v1 of
    TNil ->                                    -- 2
        v2 = TNil                              -- 3
        v3 = TNil                              -- 4
        v4 = Node v2 x v3                      -- 5
        *!tp := v4                             -- 6
        ret = ()                               -- 7
    (Node *lp *v5 *rp) ->                      -- 8
        n = *v5                                -- 9
        v6 = (x <= n)                          -- 10
        case v6 of 
        True ->                                -- 11
            v7 = (bst_insert_du x !lp) !tp     -- 12
            ret = v7                           -- 13
        False ->                               -- 14
            v8 = (bst_insert_du x !rp) !tp     -- 15
            ret = v8                           -- 16
        -- end case                            -- 17
    -- end case                                -- 18
\end{verbatim}

Here $a_0 = \{\{$\texttt{tp.[Ref.1]}$\}$,
$\{$\texttt{tp.[Ref.1,Node.2]}$\}\}$ and
$a_1 = a_0 \cup \{\{$\texttt{v1.[]}, \texttt{tp.[Ref.1]}$\}$, 
$\{$\texttt{tp.[Ref.1,Node.2]}, \texttt{v1.[Node.2]}$\}\}$.
For the \texttt{TNil} branch we remove the \texttt{v1} sharing so $a_4 =
a_3 = a_2 = a_0$ and $a_5 = a_4 \cup \{\{$\texttt{v4.[]}$\}$,
$\{$\texttt{v4.[Node.2]}$\}\}$.
After the destructive update, $a_6 = a_5 \cup \{\{$\texttt{v4.[]},
\texttt{tp.[Ref.1]}$\}$, 
$\{$\texttt{v4.[Node.2]},
\texttt{tp.[Ref.1,Node.2]}$\}\}$ (\texttt{v4} is dead and can be
removed) and $a_7 = a_6$.
For the \texttt{Node} branch we have $a_8 = a_1 \cup
\{\{$\texttt{v1.[]}, \texttt{tp.[Ref.1]}, \texttt{lp.[Ref.1]},
\texttt{rp.[Ref.1]}$\}$,
$\{$\texttt{tp.[Ref.1,Node.2]}, \texttt{lp.[Ref.1,Node.2]},
\texttt{rp.[Ref.1,Node.2]}, \texttt{v5.[Ref.1]}, \texttt{v1.[Node.2]}$\}\}$.
The same set is retained for $a_9 \ldots a_{17}$ (assuming the dead
variable \texttt{v5} is retained), the preconditions
of the function calls are satisfied and the required annotations are
present.  Finally, $a_{18} = a_{17} \cup
a_7$, which contains all the sharing for \texttt{tp},
and after eliminating local variables we get the postcondition,
which is the same as the precondition.

\section{Discussion}
\label{sec-disc}

Imprecision in the analysis of mutable parameters could potentially be
reduced by allowing the user to declare that only certain parts of a
data structure are mutable, as suggested in \cite{pawns}.
It is inevitable we lose some precision with recursion in types, but
it seems that some loss of precision could be avoided relatively easily.
The use of the empty path to represent sub-components of recursive
types results in imprecision when references are created.  For example,
the analysis of \texttt{*vp = Nil; v = *vp} concludes that the empty
component of \texttt{v} may alias with itself and the \texttt{Ref}
component of \texttt{vp} (in reality, \texttt{v} has no sharing). Instead
of the empty path, a dummy path of length one could be used.
Flagging data structures which are known to be acyclic could also
improve precision for \texttt{Case}.  A more
aggressive approach would be to unfold the recursion an extra level, at
least for some types.  This could allow us to express (non-)sharing of
separate subtrees and whether data structures are cyclic, at the cost
of more variable components, more complex pre- and post-conditions and
more complex analysis for \texttt{Assign} and \texttt{Case}.

Increasing the number of variable components also decreases efficiency.
The algorithmic complexity is affected by the representation of alias
sets.  Currently we use a naive implementation, using just ordered
pairs of variable components as the set elements and a set library
which uses an ordered binary tree.  The size of the set can be $O(N^2)$,
where $N$ is the maximum number of live variable components of the same
type at any program point (each such variable component can alias with
all the others).  In typical code the number of live variables at any
point is not particularly large.  If the size of alias sets does become
problematic, a more refined set representation could be used, such as the
set of sets of pairs representation we used in Section \ref{sec-example},
where sets of components that all alias with each other are optimised.
There are also simpler opportunities for efficiency gains, such as
avoiding sharing analysis for entirely pure code.
We have not stress tested our implementation or run substantial
benchmarks as it is intended to be a
prototype, but performance has been encouraging.
Translating the tree insertion code plus a test harness to C, which
includes the sharing analysis, takes less time than compiling
the resulting C code using GCC.  Total compilation time is less than half
that of GHC for equivalent Haskell code and less than one tenth that of
MLton for equivalent ML code.  The Pawns executable is around 3--4 times
as fast as the others.

\section{Related work}
\label{sec-related}

Related programming languages are discussed in \cite{pawns}; here we
restrict attention to work related to the sharing analysis algorithm.
The most closely related work is that done in the compiler for
Mars \cite{mars}, which extends similar work done for Mercury
\cite{mercuryCTGC} and earlier for Prolog \cite{mulkers}.  All use a
similar abstract domain based on the type folding method first proposed
in \cite{type_fold}.  Our abstract domain is somewhat more precise due
to inclusion of self-aliasing, and we have no sharing for constants.
In Mars it is assumed that constants other than numbers can share.
Thus for code such as \texttt{xs = []; ys = xs} our analysis concludes
there is no sharing between \texttt{xs} and \texttt{ys} whereas the Mars
analysis concludes there may be sharing.

One important distinction is that in Pawns sharing (and mutability) is
declared in type signatures of functions so the Pawns compiler just has to
check the declarations are consistent, rather than infer all sharing from
the code.  However, it does have the added complication of destructive
update.  As well as having to deal with the assignment primitive, it
complicates handling of function calls and case statements (the latter
due to the potential for cyclic structures).
Mars, Mercury and Prolog are essentially declarative languages.
Although Mars has assignment statements the semantics is that values are
copied rather than destructively updated --- the variable being assigned
is modified but other variables remain unchanged.  Sharing analysis
is used in these languages to make the implementation more efficient.
For example, the Mars compiler can often emit code to destructively update
rather than copy a data structure because sharing analysis reveals no
other live variables share it.  In Mercury and Prolog the analysis can
reveal when heap-allocated data is no longer used, so the code can reuse
or reclaim it directly instead of invoking a garbage collector.

These sharing inference systems use an explicit graph representation
of the sharing behaviour of each segment of code.  For example, code
$s_1$ may cause aliasing between (a component of) variables \texttt{a} and
\texttt{b} (which is represented as an edge between nodes \texttt{a}
and \texttt{b}) and between \texttt{c} and \texttt{d} and code $s_2$ may
cause aliasing between \texttt{b} and \texttt{c} and between \texttt{d} and
\texttt{e}.  To compute the sharing for the sequence $s_1$\texttt{;}$s_2$
they use the ``alternating closure'' of the sharing for $s_1$ and $s_2$,
which constructs paths with edges alternating from $s_1$ and $s_2$, for
example \texttt{a-b} (from $s_1$), \texttt{b-c} (from $s_2$), \texttt{c-d}
(from $s_1$) and \texttt{d-e} (from $s_2$).

The sharing behaviour of functions in Pawns is represented explicitly,
by a pre- and post-condition and set of mutable arguments but there is
no explicit representation for sharing of statements.  The (curried)
function \texttt{alias s} represents the sharing behaviour of \texttt{s}
and the sharing behaviour of a sequence of statements is represented by
the composition of functions.  This representation has the advantage
that the function can easily use information about the current sharing,
including self-aliases, and remove some if appropriate.
For example, in the \texttt{[]} branch of
the case in the code below the sharing for \texttt{xs} is removed and
we can conclude the returned value does not share with the argument.

\begin{verbatim}
map_const_1 :: [t] -> [Int]
    sharing map_const_1 xs = ys pre nosharing post nosharing
map_const_1 xs =
    case xs of
        [] -> xs    -- can look like result shares with xs
        (_:xs1) -> 1:(map_const_1 xs1)
\end{verbatim}

There is also substantial work on sharing analysis for logic programming
languages using other abstract domains, notably the set-sharing domain
of \cite{jl} (a set of sets of variables), generally with various
enhancements --- see \cite{bagnara} for a good summary and evaluation.
Applications include avoiding the ``occurs check'' in unification
\cite{harald} and exploiting parallelism of independent sub-computations
\cite{manuel}.  These approaches are aimed at identifying sharing of logic
variables rather than sharing of data structures.  For example, although
the two Prolog goals \texttt{p(X)} and \texttt{q(X)} share \texttt{X},
they are considered independent if \texttt{X} is instantiated to a data
structure that is ground (contains no logic variables).  Ground data
structures in Prolog are read-only and cause no problem for parallelism
or the occurs check, whether they are shared or not.
The set-sharing domain is often augmented with extra information related
to freeness (free means uninstantiated), linearity (linear means
there are no repeated occurrences of any variable)
and/or groundness \cite{bagnara}.  In Pawns there are no logic variables
but data structures are mutable, hence their sharing is important.

However, the set-sharing domain (with enhancements) has been adapted
to analysis of sharing of data structures in object oriented languages
such as Java \cite{mario}.  One important distinction is that Pawns
directly
supports algebraic data types which allow a ``sum of products'': there
can be a choice of several data constructors (a sum), where each one
consists of several values as arguments (a product).  In Java and most
other imperative and object oriented languages additional coding is
generally required to support such data types.  Products are
supported by objects containing several values but the only choice
(sum) supported directly
is whether the object is null or not.  Java objects and pointers
in most imperative languages are similar to a \texttt{Maybe} algebraic
data type, with \texttt{Nothing} corresponding to null.  A \texttt{Ref}
cannot be null.  The abstract domain of \cite{mario} uses set-sharing
plus additional information about what objects are definitely not null.
For Pawns code that uses \texttt{Ref}s this information is given
by the data type --- the more expressive types allow us to trivially
infer some information that is obscured in other languages.  For code
that uses \texttt{Maybe}, our domain can express the fact that a
variable is definitely \texttt{Nothing} by not having a self-alias of
the \texttt{Just} component.  The rich structural information in our
domain fits particularly well with algebraic data types.
There are also other approaches to and
uses of alias analysis for imperative languages, such as \cite{Landi92}
and \cite{Emami94}, but these are not aimed at precisely capturing
information about dynamically allocated data.  A more detailed discussion
of such approaches is given in \cite{mars}.

\section{Conclusion}
\label{sec-conc}

Purely declarative languages have the advantage of avoiding side effects,
such as destructive update of function arguments. This makes it easier
to combine program components, but some algorithms are hard to code
efficiently without flexible use of destructive update.  A function can
behave in a purely declarative way if destructive update is allowed,
but restricted to data structures that are created inside the function.
The Pawns language uses this idea to support flexible destructive update
encapsulated in a declarative interface.  It is designed to make all
side effects ``obvious'' from the source code.  Because there can be
sharing between the representations of different arguments of a function,
local variables and the value returned, sharing analysis is an essential
component of the compiler.  It is also used to ensure
``preservation'' of types in computations.
Sharing analysis has been used in other
languages to improve efficiency and to give some feedback to programmers
but we use it to support important features of the programming language.

The algorithm operates on (heap allocated) algebraic data types,
including arrays and closures.  In common with other sharing analysis
used in declarative languages it supports binding of variables,
construction and deconstruction (combined with selection or ``case'')
and function/procedure calls.  In addition, it supports explicit pointers,
destructive update via pointers, creation and application of closures and
pre- and post-conditions concerning sharing attached to type signatures
of functions.  It also uses an abstract domain with additional features
to improve precision.
Early indications are that the performance is acceptable: compared with
other compilers for declarative languages, the prototype
Pawns compiler supports encapsulated destructive update, is fast and
produces fast executables.

\section*{Acknowledgements}
Feedback from reviewers, particularly Gianluca Amato, was very
helpful in ironing out some important bugs in the algorithm and
improving the presentation of this paper.

\bibliography{all}
\end{document}